\documentclass[%
 reprint,
superscriptaddress,
 amsmath,amssymb,
 aps,
]{revtex4-2}

\usepackage{graphicx}% Include figure files
\usepackage{dcolumn}% Align table columns on decimal point
\usepackage{bm}% bold math
\usepackage{aas_macros}
\usepackage{color}
\usepackage{amsmath}
\usepackage{natbib}

\usepackage{setspace}

\newcommand{\average}[1]{\ensuremath{\langle#1\rangle}}

\begin{document}

\preprint{APS/123-QED}

\title{
Neutrino-antineutrino oscillations induced by strong magnetic fields in dense matter
}
% Force line breaks with \\

%\affiliation{%

\author{Hirokazu Sasaki}
\email{hsasaki@lanl.gov}
\affiliation{Division of Science, National Astronomical Observatory of Japan, \\
2-21-1 Osawa, Mitaka, Tokyo 181-8588, Japan}

\author{Tomoya Takiwaki}
\email{takiwaki.tomoya.astro@gmail.com, orcid: 0000-0003-0304-9283}
\affiliation{%
Division of Science, National Astronomical Observatory of Japan, \\
2-21-1 Osawa, Mitaka, Tokyo 181-8588, Japan}


%\affiliation{
%\affiliation{
%\affiliation{%


\date{\today}% It is always \today, today,

\begin{abstract}
 
%We simulate neutrino-antineutrino oscillations caused by strong magnetic fields in dense astrophysical sites. With the strong magnetic fields and large neutrino magnetic moments,
%Majorana neutrinos can reach flavor equilibrium. We find that the flavor equilibration during neutrino-antineutrino oscillations is sensitive to the values of the baryon density and the electron fraction  $Y_{e}$ inside the matter. The neutrino-antineutrino oscillations are suppressed in the case of the large baryon density and the finite value of $|2Y_{e}-1|$. On the other hand, the flavor equilibration occurs at $Y_{e}\sim0.5$ irrespective of the value of the baryon density. From the simulations, we propose a necessary condition of a magnetic field and the neutrino magnetic moment $\mu_{\nu}$ for the equilibration of neutrino-antineutrino  oscillations in dense matter. We also study whether such necessary condition is satisfied near the proto-neutron star by using results of neutrino hydrodynamic simulations of core-collapse supernovae. In our explosion model, the 
%flavor equilibration would be possible in the case of $\mu_{\nu}=10^{-12}\mu_{B}$ when the magnetic field on the surface of the proto-neutron star is larger than $10^{14}$ G which is the typical value of the magnetic fields of magnetars.

We simulate neutrino-antineutrino oscillations caused by strong magnetic fields in dense matter. With the strong magnetic fields and large neutrino magnetic moments,
Majorana neutrinos can reach flavor equilibrium. We find that the flavor equilibration of neutrino-antineutrino oscillations is sensitive to the values of the baryon density and the electron fraction inside the matter. The neutrino-antineutrino oscillations are suppressed in the case of the large baryon density in neutron (proton)-rich matter. On the other hand, the flavor equilibration occurs when the electron fraction is close to $0.5$ even in the large baryon density. From the simulations, we propose a necessary condition for the equilibration of neutrino-antineutrino  oscillations in dense matter. We also study whether such necessary condition is satisfied near the proto-neutron star by using results of neutrino hydrodynamic simulations of core-collapse supernovae. In our explosion model, the 
flavor equilibration would be possible if the magnetic field on the surface of the proto-neutron star is larger than $10^{14}$ G which is the typical value of the magnetic fields of magnetars.

\end{abstract}

\maketitle

%\tableofcontents

\section{Introduction}

Neutrinos are produced through weak interactions in various explosive astrophysical sites \cite{Vitagliano:2019yzm}. 
The detection of neutrino bursts ($\sim20$ events) from Supernova 1987A has opened up the possibility of identifying explosive dynamics and properties of particle physics from neutrino observations \cite{Kamiokande:1987A,IMB:1987A,Baksan:1987A}. Current operational neutrino observatories can detect $\sim 10^{4}$-$10^{6}$ events during a neutrino burst from a supernova in our galaxy (see, e.g., reviews \cite{Horiuchi2018WhatDetection,Janka2017NeutrinoSupernovae,Hix2016TheSupernovae,Muller2016TheModels,Mirizzi2015SupernovaDetection,Foglizzo2015TheExperiments,Burrows2013ColloquiumTheory,Kotake2012Core-collapseRelativity}). High statistical neutrino signals in neutrino detectors help investigate the detailed mechanism of the explosion and behaviors of neutrino oscillations inside the progenitor star. 

Neutrinos propagating inside astrophysical sites are affected by neutrino coherent forward scatterings with background particles. Charged current interactions of $\nu_{e}(\bar{\nu}_{e})$ with background electrons induce significant flavor conversions called ``Mikheyev- Smirnov-Wolfenstein (MSW) effect" \cite{Wolfenstein1978NeutrinoMatter,Mikheev:1986gs} when the number density of electrons inside the star decreases down to a critical density. Neutrino-neutrino interactions in dense neutrino gas cause a self-refraction term in the neutrino Hamiltonian \cite{Fuller1987,Pantaleone:1992eq,Pantaleone:1992xh,Sigl:1992fn,McKellar:1992ja,Yamada:2000za,Balantekin:2006tg,Cardall:2007zw,Pehlivan:2011hp,Vlasenko:2013fja,Volpe:2013jgr,Blaschke:2016xxt,Birol:2018qhx,Richers:2019grc}. It is believed that, in core-collapse supernovae (CCSNe), the nonlinear potential of the self-refraction effect induces ``collective neutrino oscillations" (CNO) outside a proto-neutron star (PNS) (see, e.g., a review \cite{Duan:2009cd}). Earlier numerical studies of CNO \cite{Duan:2006jv,Duan:2006an,Fogli:2007bk,Raffelt:2007xt,Dasgupta:2007ws,Dasgupta:2009mg,Dasgupta:2010cd,Duan:2010bf,Friedland:2010sc,Mirizzi:2010uz} found the so-called ``spectral splits/swap" phenomenon, which exchanges neutrino spectra around certain critical energies. In the last decade, ``multiangle calculations" were carried out by employing results of neutrino radiation hydrodynamics of CCSNe \cite{Cherry:2010yc,Cherry:2011fm,Chakraborty:2011nf,Chakraborty:2011gd,Wu:2014kaa,Sasaki:2017jry,Zaizen:2018wfg,Sasaki:2019jny,Zaizen2020JCAP,Zaizen:2020xum}. In such multiangle simulations, the CNO increases energetic $\nu_{e}$ and $\bar{\nu}_{e}$ even though the ``matter suppression" \cite{EstebanPretel:2008ni,Mirizzi:2010uz,Saviano:2012yh,Chakraborty:2011gd} smears the spectral splits in the neutrino distribution. The enhanced spectra of $\nu_{e}$ and $\bar{\nu}_{e}$ potentially affect neutrino detections \cite{Wu:2014kaa,Sasaki:2019jny,Zaizen2020JCAP,Zaizen:2020xum} and nucleosynthesis inside the star such as $\nu p$-process \cite{MartinezPinedo:2011br,Pllumbi:2014saa,Sasaki:2017jry,Xiong:2020ntn} and  $\nu$-process \cite{Ko:2020rjq}. In neutron star mergers, neutrino-neutrino interactions can induce the ``matter neutrino resonances" \cite{Malkus:2014iqa,Malkus:2015mda,Wu:2015fga,Zhu:2016mwa,Frensel:2016fge,Chatelain:2016xva,Tian:2017xbr,Vlasenko:2018irq,Shalgar:2017pzd} and affect rapid neutron capture process ($r$-process) nucleosynthesis \cite{Malkus:2015mda,Wu:2017drk,George:2020veu}. Furthermore, the possibility of fast-pairwise collective neutrino oscillations which may occur in 
the scale of $\sim\mathcal{O}(10^{-5})\ \mathrm{km}$ are studied in CCSNe, neutron star mergers, and the early universe (see e.g. a review \cite{Tamborra:2020cul}).

The finite neutrino magnetic moment can lead to flavor conversions between left-handed neutrinos and right-handed (anti)neutrinos in strong magnetic fields. Such a magnetic field effect results in conversions between active and sterile neutrinos in the case of Dirac neutrinos and neutrino-antineutrino oscillations in the case of Majorana neutrinos. The neutrino magnetic moment is a key to investigate new physics beyond the standard model (see, e.g., reviews \cite{Balantekin:2018,Giunti:2015}). The value of neutrino magnetic moment is constrained from recent neutrino experiments, e.g., GEMMA \cite{Beda:2013mta}, Borexino \cite{Borexino:2017fbd}, XMASS-I \cite{Abe:2020nwr}, and XENON1T \cite{Aprile:2020tmw}. Among such neutrino experiments, Borexino experiment provides a stringent upper limit on the neutrino magnetic moment: $\mu_{\nu}<2.8\times10^{-11}\mu_{B}$ ($90\%$ C.L) \cite{Borexino:2017fbd}, where $\mu_{B}$ is the Bohr magneton. The value of the neutrino magnetic moment can be also constrained from neutrino energy loss in globular clusters \cite{Arceo-Diaz:2015pva,Capozzi:2020} and intermediate-mass stars \cite{Mori:2020niw}. Currently the most stringent upper limit on the neutrino magnetic moment is $\mu_{\nu}<1.2\times10^{-12}\mu_{B}$ \cite{Capozzi:2020}. In the case of Majorana neutrinos, the resonant neutrino-antineutrino conversions called ``resonant spin-flavor”(RSF) conversions are studied in CCSNe \cite{Lim:1987tk,Akhmedov:1992ea,Akhmedov:1993sh,Totani:1996wf,Nunokawa:1996gp,Ando:2002sk,Ando:2003pj,Ando:2003is,Akhmedov:2003fu,Ahriche:2003wt,Yoshida:2009ec}. Such resonant flavor conversions are induced by the finite neutrino magnetic moment in strong magnetic fields, and significant $\nu-\bar{\nu}$ transitions occur at the resonance baryon densities (see e.g. \cite{Ando:2003pj}). The RSF conversions are sensitive to the sign of $2Y_{e}-1$ \cite{Yoshida:2009ec} where the $Y_{e}$ is the electron fraction of the supernova material.

The neutrino-antineutrino oscillations considering neutrino-neutrino interactions are studied in Refs.\cite{Dvornikov:2011dv,deGouvea:2012hg,deGouvea:2013zp,Abbar:2020ggq,Yuan2021}. The flavor equilibration of neutrino-antineutrino oscillations occurs in the scale determined by a neutrino magnetic potential in strong magnetic fields \cite{Abbar:2020ggq}. Such equilibration phenomenon is different from the RSF conversion and suppressed by strong neutrino-neutrino interactions. However, the role of matter potentials on the neutrino-antineutrino oscillations is still unknown. The matter suppression, as confirmed in multiangle calculations, should be important in dense background matter. 

In this work, we study the effect of matter suppression on equilibrations of neutrino-antineutrino oscillations caused by magnetic fields and discuss the possibility of such curious oscillations in astrophysical sites. In Sec.\ref{sec:simulation}, we carry out numerical simulations of the neutrino-antineutrino oscillations by assuming strong magnetic field ($B\gtrsim10^{14}$G) and typical values of neutrino potentials in CCSNe and neutron-star mergers. From the simulations, we reveal the mechanism of matter suppression on the neutrino-antineutrino oscillations and propose a necessary condition of a magnetic potential to realize the equilibration of neutrino-antineutrino oscillations in dense matter. In Sec.\ref{sec:check magnetic effect in supernovae}, we verify that the necessary condition of the neutrino-antineutrino oscillations given in Sec.\ref{sec:simulation} is satisfied outside a PNS in CCSNe based on supernova hydrodynamic simulations. Finally, our results are summarized in Sec.\ref{sec:Summary and Discussion}.\\

\section{ 
Simulations of flavor equilibration in magnetic fields
}\label{sec:simulation}

In the case of Majorana neutrinos, we perform simulations of the neutrino-antineutrino oscillations in magnetic fields by changing values of the MSW matter potential and the electron fraction based on a numerical setup of Ref.~\cite{Abbar:2020ggq}. Equilibrium values of diagonal components of neutrino density matrices are reproduced analytically in Appendix \ref{sec:appendix analytical explanation}.

\subsection{Equations of motion of flavor conversions}
\label{sec:numerical setup}

Here, we employ the ``neutrino line model" \cite{Duan:2014gfa,Abbar:2018beu} to simulate neutrino-antineutrino oscillations in magnetic fields. Flavor conversions of neutrinos with an emission angle $\theta$ at a radius $r$ are calculated by solving the Liouville-von Neumann equation \cite{Duan:2014gfa,Abbar:2018beu,Abbar:2020ggq}:
\begin{equation}
\label{eq:Liouville-von Neumann}
    \cos\theta\frac{\partial}{\partial r}D=-i[H,D],
\end{equation}
where $D(r,\theta)$ and $H(r,\theta)$ are $6\times6$ neutrino density matrix and Hamiltonian, respectively. The density matrix is given by 
\begin{equation}
\label{eq:neutrino density matrix 6times6}
D=\left(
\begin{array}{c c}
\rho_{\theta}&X_{\theta}\\
X^{\dagger}_{\theta}&\bar{\rho}_{\theta}\\
\end{array}\right),
\end{equation}
where diagonal components $\rho_{\theta}$ and $\bar{\rho}_{\theta}$ are $3\times3$ density matrices of neutrinos and antineutrinos, respectively \cite{Abbar:2020ggq}. The subscript, $\theta$, means $\theta$-dependence. Here, we impose normalization of neutrino density matrices: $\mathrm{tr}[\rho_{\theta}+\bar{\rho}_{\theta}]=1$ where the ``tr" represents the trace of a $3\times3$ matrix (e.g.,$\mathrm{tr}[\rho_{\theta}]=\sum_{\alpha=e,\mu,\tau}(\rho_{\theta})_{\alpha\alpha}$). The non-diagonal component $X_{\theta}$ is a correlation between $\nu$ and $\bar{\nu}$ which is usually negligible without a magnetic field. A finite value of $X_{\theta}$ induces neutrino-antineutrino oscillations. The Hamiltonian of neutrinos are decomposed into three terms:
\begin{equation}
\label{eq:Hamiltonian}
H(r,\theta)=H_{\mathrm{vac}}+H_{\mathrm{matter}}+H_{\nu\nu}.
\end{equation}
The first term on the right hand side represents the vacuum Hamiltonian in the magnetic field \cite{Ando:2002sk,Ando:2003pj,Yoshida:2009ec,deGouvea:2012hg,deGouvea:2013zp}:
\begin{equation}
\label{eq:vacuum}
H_{\mathrm{vac}}=\left(
\begin{array}{c c}
\Omega(E)&V_{\mathrm{mag}}\\
-V_{\mathrm{mag}}&\Omega^{*}(E)\\
\end{array}\right),
\end{equation}
\begin{equation}
V_{\mathrm{mag}}=B_{T}\left(
\begin{array}{c c c}
0&\mu_{e\mu}&\mu_{e\tau}\\
-\mu_{e\mu}&0&\mu_{\mu\tau}\\
-\mu_{e\tau}&-\mu_{\mu\tau}&0\\
\end{array}\right),
\end{equation}
where $\mu_{\alpha\beta}(\alpha,\beta=e,\mu,\tau)$ are neutrino magnetic moments and $B_{T}$ is a transverse component of the magnetic field perpendicular to the direction of the neutrino emission. We assume the same neutrino magnetic moment irrespective of flavor dependence: $\mu_{e\mu}=\mu_{e\tau}=\mu_{\mu\tau}=\mu_{\nu}$. The diagonal component $\Omega(E)$ in Eq.(\ref{eq:vacuum}) is $3\times3$ vacuum Hamiltonian of neutrinos without a magnetic field \cite{Sasaki:2017jry}. Here, we use the same neutrino mixing parameter set: $\{\Delta m^{2}_{21},\Delta m^{2}_{32},\theta_{12},\theta_{13},\theta_{23} ,\delta_{CP}\}$ as that of Ref.\cite{Sasaki:2017jry}. We set normal neutrino mass hierarchy($\Delta m^{2}_{32}>0$) in all of our simulations. We investigate oscillation behaviors of single energy neutrinos: $E=1$ MeV. 
The flavor equilibration phenomenon in the scale of $\sim (\mu_{\nu}B_{T})^{-1}$ does not depend on the choice of neutrino mixing parameters and neutrino energy qualitatively. The second term on the right hand side of Eq.(\ref{eq:Hamiltonian}) is the matter potential \cite{Ando:2002sk,Ando:2003pj,Yoshida:2009ec,deGouvea:2012hg,deGouvea:2013zp}. It is given by
\begin{equation}
H_{\mathrm{matter}}=\left(
\begin{array}{c c}
V_{\mathrm{matter}}&0\\
0&-V_{\mathrm{matter}}
\end{array}\right),
\end{equation}

\begin{equation}
\label{eq:matter matrix 2}
V_{\mathrm{matter}}=-\frac{\lambda_{n}}{2}I_{3\times3}+\lambda_{e}\left(
\begin{array}{c c c}
1&0&0\\
0&0&0\\
0&0&0
\end{array}\right),
\end{equation}

\begin{equation}
\lambda_{e(n)}=\sqrt{2}G_{F}n_{e(n)},
\end{equation}
where $n_{e(n)}$ is the number density of electrons(neutrons), respectively and $I_{3\times3}=\mathrm{diag}(1,1,1)$. The first term on the right hand side of Eq.(\ref{eq:matter matrix 2}) does not contribute to flavor conversions without magnetic field. On the other hand, such term plays important role in neutrino-antineutrino oscillations caused by magnetic field effect.

The third term on the right hand side of Eq.(\ref{eq:Hamiltonian}) shows the potential of neutrino-neutrino interactions which are sources of CNO. Here, we consider single energy neutrinos and focus on angular dependence of neutrinos by using the neutrino line model \cite{Duan:2014gfa,Abbar:2018beu}. The potential of neutrino-neutrino interactions in this model is written as 
\begin{equation}
H_{\nu\nu}=\left(
\begin{array}{c c}
V_{\nu\nu}&0\\
0&-V^{*}_{\nu\nu}
\end{array}\right),
\end{equation}
\begin{equation}
\label{eq:self interaction matrix 2}
\begin{split}
V_{\nu\nu}&=\sqrt{2}G_{F}n_{\nu}\int^{\theta_{\mathrm{max}}}_{-\theta_{\mathrm{max}}}\mathrm{d}\theta^{\prime}\ \left(
1-\cos(\theta-\theta^{\prime})
\right)\\
&\times\left\{
(\mathrm{tr}[\rho_{\theta^{\prime}}]-\mathrm{tr}[\bar{\rho}^{*}_{\theta^{\prime}}])I_{3\times3}+(\rho_{\theta^{\prime}}-\bar{\rho}^{*}_{\theta^{\prime}})
\right\},
\end{split}
\end{equation}

\begin{equation}
n_{\nu}=\sum_{\alpha=e,\mu,\tau}(n_{\nu_{\alpha}}+n_{\bar{\nu}_{\alpha}}),
\end{equation}
where $\theta_{\mathrm{max}}$ is the maximum neutrino emission angle in the line model \cite{Abbar:2020ggq}. The $n_{\nu}$ is a summation of initial number densities of all species of neutrinos. The initial values of diagonal terms of neutrino density matrices are given by $(\rho_{\theta})_{\alpha\alpha}=n_{\nu_{\alpha}}/n_{\nu}, (\bar{\rho}_{\theta})_{\alpha\alpha}=n_{\bar{\nu}_{\alpha}}/n_{\nu}(\alpha=e,\mu,\tau)$. \\

\subsection{
Flavor equilibrations in different $(\lambda_{e},Y_{e}$)}
\label{sec:simulation result}

We perform simulations of neutrino-antineutrino oscillations based on numerical setup in Sec.\ref{sec:numerical setup}. The initial neutrino number density is given by $n_{\bar{\nu}_{e}}/n_{\nu_{e}}=0.7$ and  $n_{\nu_{x}}/n_{\nu_{e}}=n_{\bar{\nu}_{x}}/n_{\nu_{e}}=0.4(x=\mu,\tau)$ as done in Ref.\cite{Abbar:2020ggq}. In addition, we fix the strength of the nonlinear potential and that of magnetic potential as $\sqrt{2}G_{F}n_{\nu_{e}}=10^{-2}\,\mathrm{cm}^{-1}$ and $\Omega_{\mathrm{mag}}=\mu_{\nu}B_{T}=0.1\,\mathrm{cm}^{-1}$. The strength of the matter potential $\lambda_{e}$ and the value of electron fraction $Y_{e}$ are variables for our simulations. The corresponding baryon density to a given $\lambda_{e}$ is $\rho_{b}=5\times10^{7}\,\mathrm{g}/\mathrm{cm}^{3}(Y_{e}/0.5)^{-1}(\lambda_{e}/0.1\,\mathrm{cm})$. Non-diagonal components in Eq.(\ref{eq:neutrino density matrix 6times6}) are set to zero at $r=0$. In our simulations, we set a maximum emission angle to $\theta_{\mathrm{max}}=\pi/3$.

\begin{figure}[t]
\includegraphics[width=0.95\linewidth]{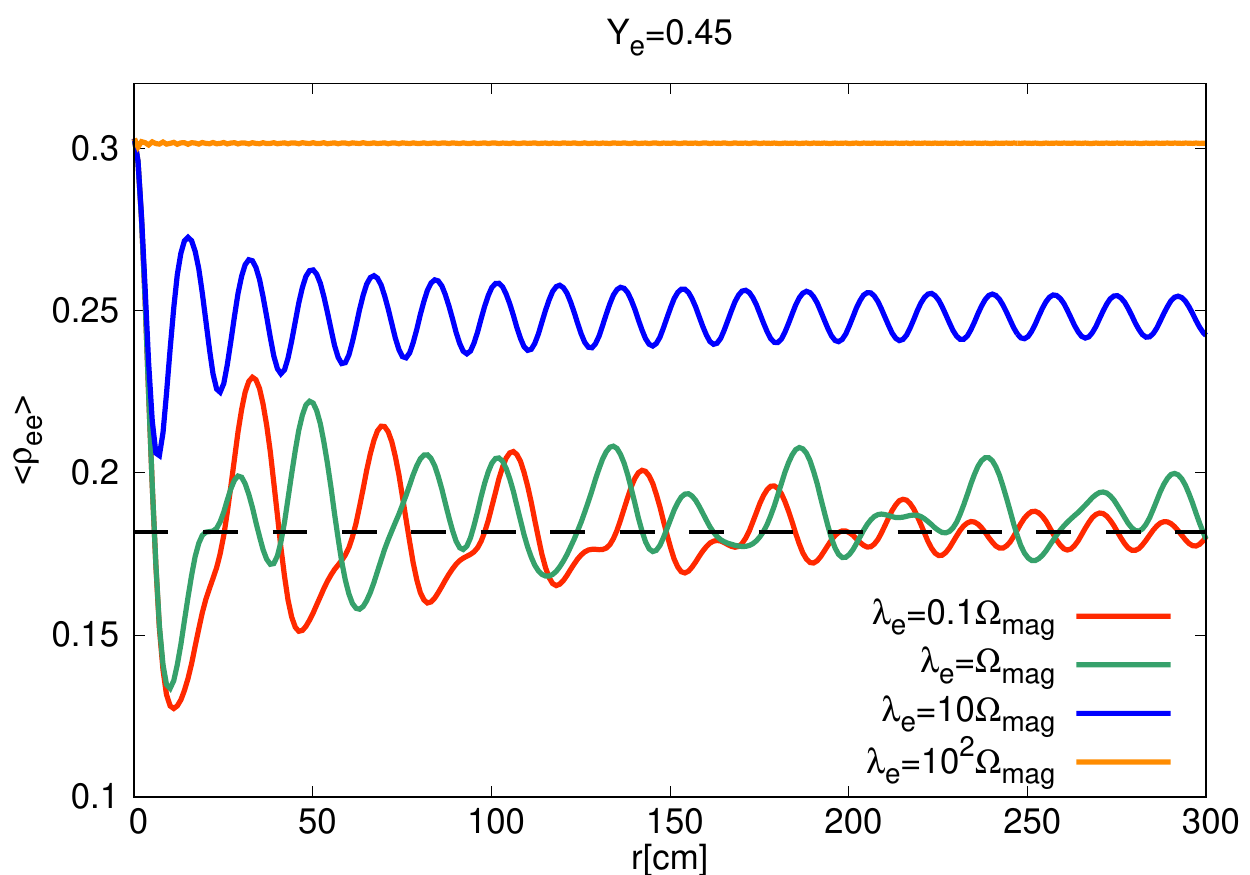}\\
\includegraphics[width=0.95\linewidth]{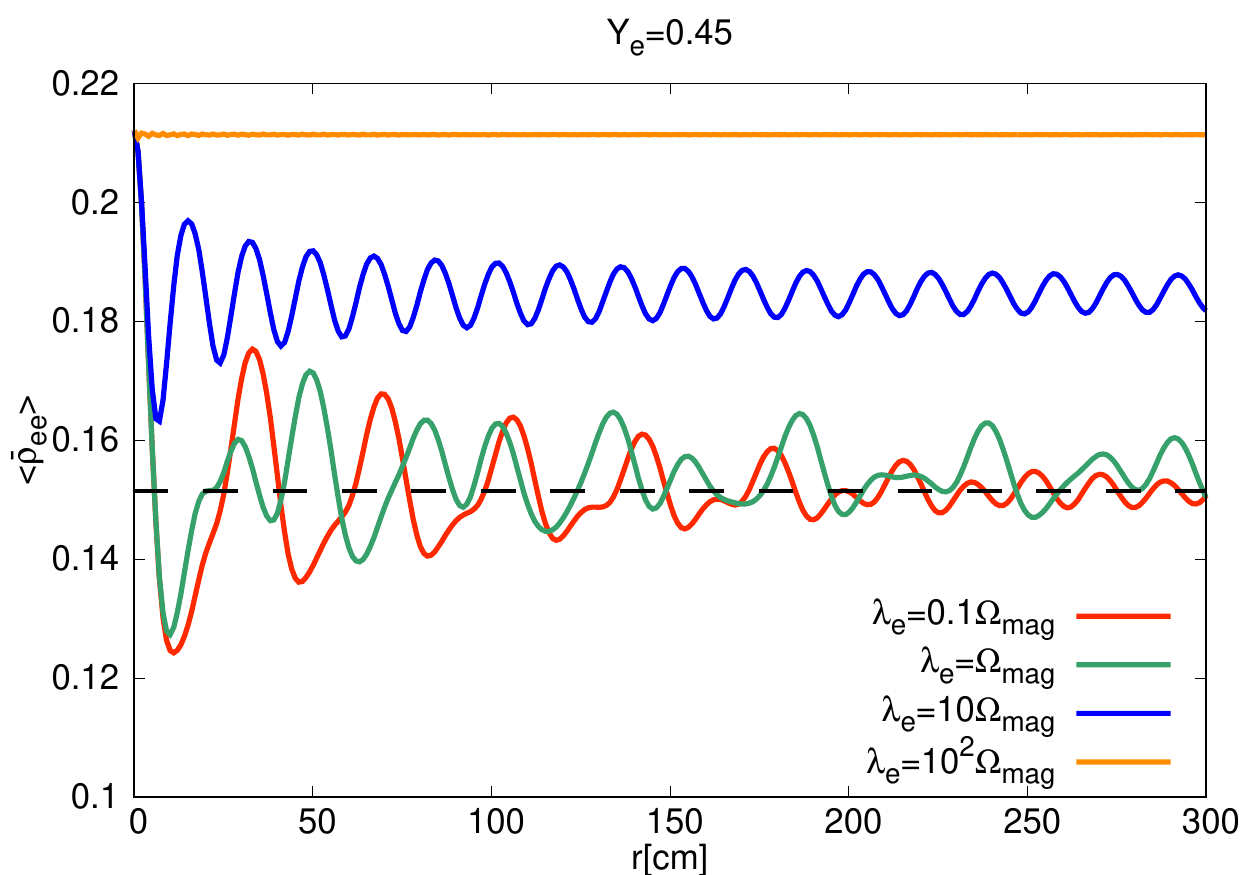}
\caption{%%
The top panel shows evolution of $\average{\rho_{ee}}$ as defined in Eq.(\ref{eq:angle average rho_e}) in different values of matter potential $\lambda_{e}$ with a constant value of electron fraction $Y_{e}=0.45$. The value of a black dash line in the top panel corresponds to $(n_{\nu_{e}}+2n_{\nu_{x}})/3n_{\nu}=2/11$ which represents an equilibrium value of $\average{\rho_{ee}}$. The bottom panel shows the case of $\average{\bar{\rho}_{ee}}$ in Eq.(\ref{eq:angle average rho_eb}). The value of black dash line in the bottom panel is $(n_{\bar{\nu}_{e}}+2n_{\nu_{x}})/3n_{\nu}=5/33$.
}
\label{fig:matter_dep}
\end{figure}

The top (bottom) panel of Fig.\ref{fig:matter_dep} shows evolution of angle averaged ratios of $\nu_{e}$ ($\bar{\nu}_{e}$), respectively in a fixed value of the electron fraction $Y_{e}=0.45$ and different values of $\lambda_{e}$. Such angular average diagonal components are defined by
\begin{equation}
\label{eq:angle average rho_e}
\average{\rho_{ee}}=\frac{3}{2\pi}\int^{\frac{\pi}{3}}_{-\frac{\pi}{3}}\mathrm{d}\theta\ \rho_{ee,\theta},
\end{equation}
\begin{equation}
\label{eq:angle average rho_eb}
\average{\bar{\rho}_{ee}}=\frac{3}{2\pi}\int^{\frac{\pi}{3}}_{-\frac{\pi}{3}}\mathrm{d}\theta\ \bar{\rho}_{ee,\theta}.
\end{equation}

The red and green curves in Fig.\ref{fig:matter_dep} represent the equilibration of neutrino-antineutrino oscillations when the potential of the magnetic field is no smaller than the matter potential: $\Omega_{\mathrm{mag}}\geq\lambda_{e}$. Such results are consistent with the equilibrium of three flavor neutrinos shown in Ref.\cite{Abbar:2020ggq}. The wave length of the neutrino-antineutrino oscillations is the order of $\sim\Omega_{\mathrm{mag}}^{-1}=10$\,cm. In such a strong magnetic field, all flavors of neutrinos and antineutrinos couple with flavor conversions. The black dash lines in the top and the bottom panels of Fig.\ref{fig:matter_dep} represent equilibrium values of three flavor conversions. Such equilibrium values of three flavor conversions are reproduced in Sec.\ref{sec:appendix three flavor}, analytically. In the strong magnetic field, $\Omega_{\mathrm{mag}}$ becomes dominant in the neutrino Hamiltonian. The large $\Omega_{\mathrm{mag}}$ plays important role to increase an amplitude of a correlation matrix $X_{\theta}$ in Eq.(\ref{eq:neutrino density matrix 6times6}). Especially, a finite value of $X_{xy}$ ($X_{yx}$) which shows a correlation between $\nu_{x}$ ($\nu_{y}$) and $\bar{\nu}_{y}$ ($\bar{\nu}_{x}$), respectively, enables active three flavor neutrino-antieneutrino oscillations. Here, flavors $x$ and $y$ are given by rotation of flavor basis \cite{Dasgupta:2007ws}. The detail mechanism of such three flavor oscillations in the magnetic field is shown in Sec.\ref{sec:appendix three flavor}.

In the case of $\lambda_{e}=10\Omega_{\mathrm{mag}}$ (blue lines in Fig.\ref{fig:matter_dep}), on the other hand, the neutrino-antineutrino oscillations occur but such flavor conversions are suppressed because of the large matter potential. The flavor conversions do not reach the equilibrium values (black dash lines in in Fig.\ref{fig:matter_dep}). In the case of more dense matter $\lambda_{e}=10^{2}\Omega_{\mathrm{mag}}$ (dark orange lines in Fig.\ref{fig:matter_dep}), values of $\average{\bar{\rho}_{ee}}$ and $\average{\rho_{ee}}$ are constant irrespective of the radius, so that the flavor conversions are completely suppressed. Such results indicate that we need to take into account the contributions of matter potential in order to study the neutrino-antineutrino oscillations in the strong magnetic fields of explosive astrophysical sites. The mechanism of matter suppression as shown in Fig.\ref{fig:matter_dep} is discussed in Sec.\ref{sec:complete matter suppression}. Here, we set a constant value of $Y_{e}=0.45$, so that the matter potential $|\lambda_{e}-\lambda_{n}|\propto|2Y_{e}-1|$ becomes finite and contributes to the matter suppression. However, the matter potential is sensitive to the value of $Y_{e}$, so that matter suppression should also depend on the value of $Y_{e}$.

\begin{figure}[t]
\includegraphics[width=0.95\linewidth]{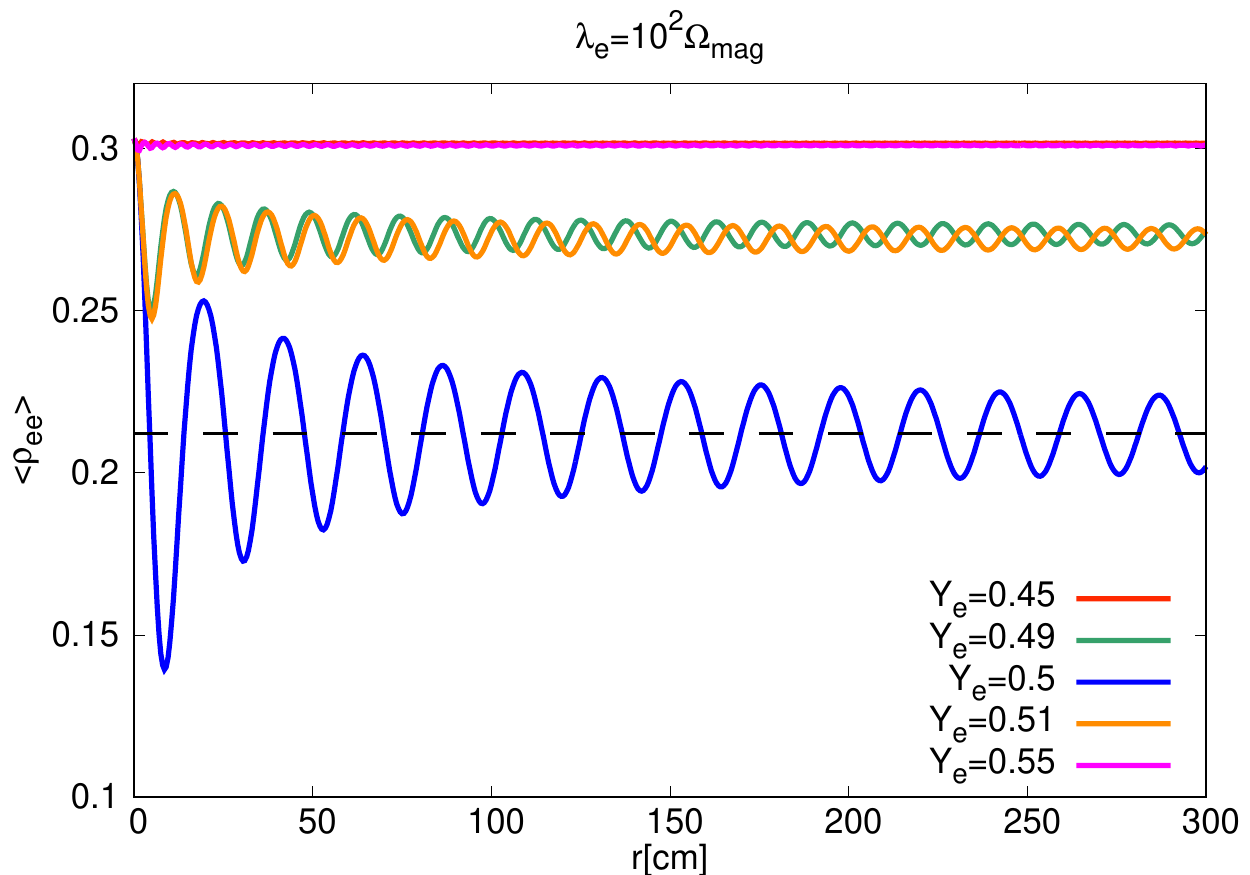}\\
\includegraphics[width=0.95\linewidth]{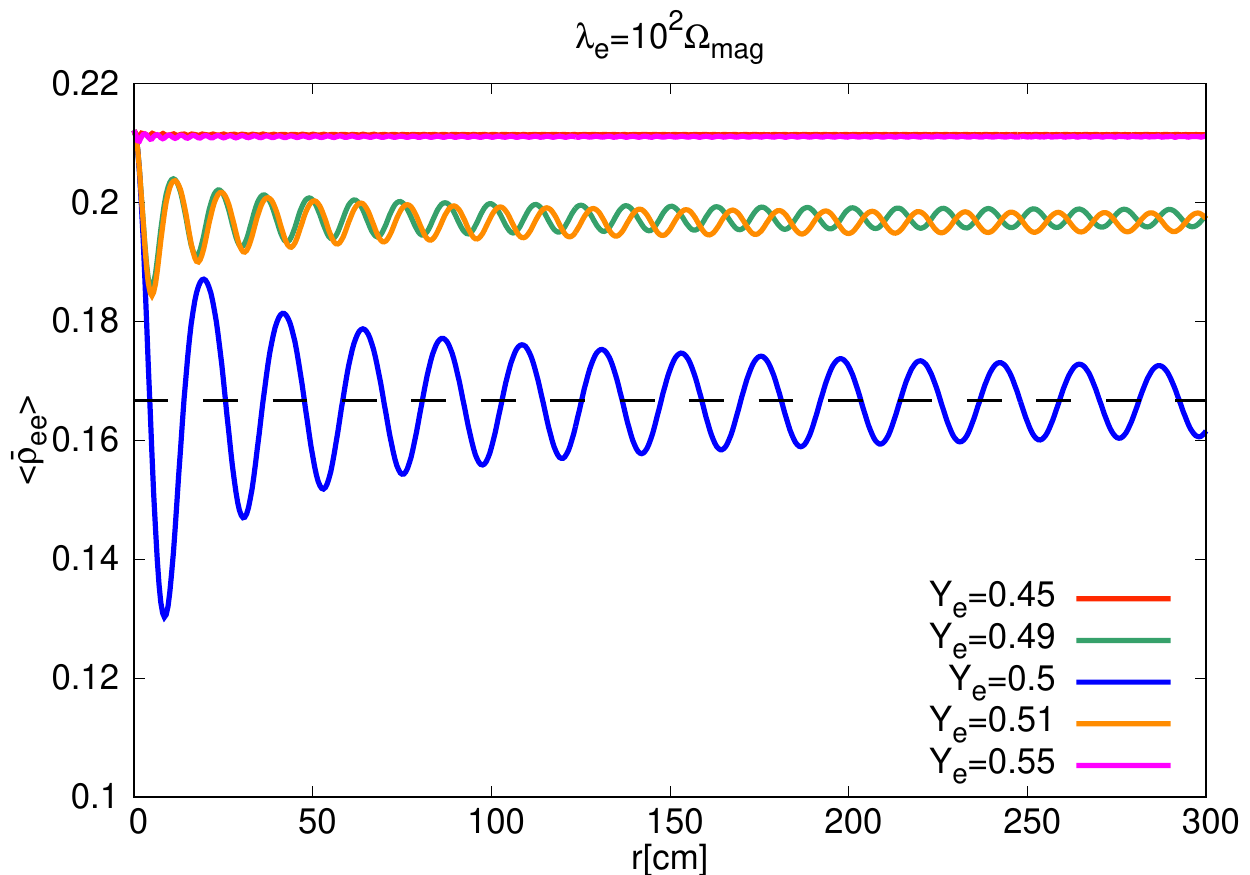}
\caption{%%
The neutrino-antineutrino oscillations as shown in Fig.\ref{fig:matter_dep} in different values of electron fraction $Y_{e}$ around $0.5$. The strength of MSW matter potential is fixed by $\lambda_{e}=10^{2}\Omega_{\mathrm{mag}}$. The black dash line in the top panel corresponds to $(n_{\nu_{e}}+n_{\nu_{x}})/2n_{\nu}=7/33$. The black dash line in the bottom panel shows $(n_{\bar{\nu}_{e}}+n_{\nu_{x}})/2n_{\nu}=1/6$.
}
\label{fig:ye_dep}
\end{figure}

Fig.\ref{fig:ye_dep} shows dependence of electron fraction $Y_{e}$ in the neutrino-antineutrino oscillations. The MSW matter potential is fixed by $\lambda_{e}=10^{2}\Omega_{\mathrm{mag}}$. Any flavor conversions are suppressed in the cases of $Y_{e}=0.45$ (red solid lines) and  $Y_{e}=0.55$ (magenta solid lines). On the other hand, flavor conversions become prominent as the value of $Y_{e}$ is close to $0.5$ (see green and dark-orange solid lines). The matter suppression disappears in the case of $Y_{e}=0.5$ (blue solid lines). Here, we fix the strength of matter potential as $\lambda_{e}=10^{2}\Omega_{\mathrm{mag}}$, but flavor conversions at $Y_{e}=0.5$ also appear even though more dense matter potential such as $\lambda_{e}=10^{3}\Omega_{\mathrm{mag}}$ is employed. The matter suppression disappears at $Y_{e}=0.5$ because matter potentials of charged and neutral current reactions are canceled out: $\lambda_{e}=\lambda_{n}$. The black dash line in the top (bottom) panel of Fig.\ref{fig:ye_dep} shows an equilibrium value of $\average{\rho_{ee}}$ ($\average{\bar{\rho}_{ee}}$) at $Y_{e}=0.5$, respectively. These equilibrium values are different from that in Fig.\ref{fig:matter_dep} because one of flavors is decoupled from flavor conversions. The matter term in Eq.(\ref{eq:matter matrix 2}) induces two types of matter potentials such as $|\lambda_{e}-\lambda_{n}|$ and $\lambda_{n}$ in equation of motion of neutrino density matrices. The $|\lambda_{e}-\lambda_{n}|$ disappears at $Y_{e}=0.5$ but another matter potential $\lambda_{n}$ is not eliminated. The large $\lambda_{n}$ results in decoupling of $\nu_{x}$ and $\bar{\nu}_{x}$ in flavor conversions. The detail of such one flavor decoupling is shown in Sec.\ref{sec:one flavor decoupling in three flavor oscillations} and equilibrium values in Fig.\ref{fig:ye_dep} are reproduced analytically. Our simulations in a parameter space of $(\lambda_{e},Y_{e})$ indicate that the strong magnetic field potential $\Omega_{\mathrm{mag}}$: 
\begin{equation}
\label{eq:lambda}
\Omega_{\mathrm{mag}}\geq\lambda,
\end{equation}
is necessary for the equilibration of neutrino-antineutrino oscillations where we define $\lambda=|\lambda_{e}-\lambda_{n}|$. The above condition can be satisfied even in dense astrophysical sites when $Y_{e}$ in the dense matter is close to $0.5$.\\

\section{Possibility of supernova neutrino-antineutrino oscillations in magnetic fields}
\label{sec:check magnetic effect in supernovae}

The necessary condition of the neutrino-antineutrino oscillations proposed in our simulations is extended by taking account of the contribution from neutrino-neutrino interactions inside CCSNe. We investigate the possibility of the flavor equilibrium of neutrino-antineutrino oscillations in CCSNe by employing matter profiles and neutrino spectra obtained in a hydrodynamic simulation of CCSNe of a 11.2 $M_\odot$ progenitor. The case of electron capture supernova (ECSN) is shown in Appendix \ref{sec:appendix ECSN potential comparison}.

\subsection{The strength of potentials in neutrino Hamiltonian}

As suggested in Eq.(\ref{eq:lambda}), the occurrence of the neutrino-antineutrino oscillations can be evaluated by a comparison of $\Omega_{\mathrm{mag}}=\mu_{\nu}B_{T}$ with other potentials in neutrino Hamiltonian. We focus on the neutrino-antineutrino oscillations near the PNS ($r\sim \mathcal{O}(10-10^{3})$ km) where the magnetic field may be large enough to satisfy the Eq.(\ref{eq:lambda}). In order to discuss the possibility of such flavor conversions in magnetic fields, we need matter profiles and neutrino spectra of deep inside CCSNe. We employ different time snapshots of matter profiles and neutrino spectra obtained in a hydrodynamic simulation of supernova explosions.
The neutrino radiation-hydrodynamic simulations are performed by {\small 3DnSNe-IDSA} code \cite{Takiwaki2016MNRAS} (see  Ref~\cite{Matsumoto2020MNRAS} for hydrodynamic method, Refs.~\cite{Takiwaki2014ApJ,Kotake2018ApJ} for neutrino transport, and  Ref.~\cite{O'Connor2018JPhG} for code comparison). The initial condition is taken from 
$11.2\,{M_{\odot}}$ progenitor model of Woosley et al., ($2002$) \cite{Woosley2002RvMP}.
The hydrodynamic profiles of this model is also used in Ref.~\cite{Cherry:2019vkv} whose detailed setup is written in Ref.~\cite{Zaizen2020JCAP}.

Since we employ the profile that is taken from the pure hydrodynamic calculation, there is no spacial profile of the magnetic field in our explosion model. We assume a transverse component of a dipole magnetic field whose radial dependence is given by $B_{T}\propto r^{-3}$ as employed in previous works \cite{Ando:2002sk,Ando:2003pj,Yoshida:2009ec}:
\begin{equation}
\label{eq:magnetic field radial profile}
B_{T}=B_{0}\left(
\frac{R_{\nu}}{r}
\right)^{3},
\end{equation}
where $R_{\nu}$ is a radius of the neutrino sphere and $B_{0}$ is a strength of a transverse magnetic field to the direction of neutrino emission at $r=R_{\nu}$. Here, we set $R_{\nu}=30$ km and focus on neutrino emission along the radial direction from the equator of the PNS. We fix the value of the neutrino magnetic moment: $\mu_{\nu}=10^{-12}\mu_{B}$ satisfying the current upper limit of neutrino magnetic moments. The potential of the magnetic field interaction is written as $\Omega_{\mathrm{mag}}=\mu_{\nu}B_{T}$, so that the neutrino-antineutrino oscillations would be induced even in the case of smaller value of $\mu_{\nu}$ than $10^{-12}\mu_{B}$ if we set a larger magnetic field, $B_{0}$, on the surface of neutrino sphere.

Neutrino-neutrino interactions should be taken into account in neutrino oscillations near the PNS. Behaviors of the neutrino-antineutrino oscillations in a magnetic field are sensitive to the strength of neutrino-neutrino interactions. The large potential of neutrino-neutrino interactions also prevents flavor conversions if the potential of the magnetic field is much smaller than that of neutrino-neutrino interactions \cite{Abbar:2020ggq}. If the neutrino emission angle is large, the strength of neutrino-neutrino interactions can be characterized by $\sqrt{2}G_{F}|n_{\nu_{e}}-n_{\bar{\nu}_{e}}|$ without considering a contribution from an angle factor $1-\cos(\theta-\theta^{\prime})$ in Eq.(\ref{eq:self interaction matrix 2}). However, such an angle factor can not be ignored when the maximum neutrino emission angle is small. The strength of neutrino-neutrino interactions depends on the value of the maximum neutrino emission angle. We use the bulb model \cite{Duan:2006an} to determine the strength of neutrino-neutrino interactions outside a neutrino sphere. In the bulb model, the maximum neutrino emission angle at the radius $r$ is given by $\sin\theta_{\mathrm{max}}=R_{\nu}/r$. The potential of neutrino-neutrino interactions in Eq.(\ref{eq:self interaction matrix 2}) is a function of neutrino emission angle $\theta$. Such angle dependence is removed by assuming $\theta=0$ under the single angle approximation \cite{Duan:2006an}. The strength of neutrino-neutrino interactions including the angular factor is described by

\begin{equation}
\label{eq:zeta}
\begin{split}
\zeta&=\frac{\sqrt{2}G_{F}}{2\pi R_{\nu}^{2}}\left|
\frac{L_{\nu_{e}}}{\average{E_{\nu_{e}}}}-\frac{L_{\bar{\nu}_{e}}}{\average{E_{\bar{\nu}_{e}}}}
\right|\int^{1}_{\cos\theta_{\mathrm{max}}}\mathrm{d}\cos\theta\ \left(
1-\cos\theta
\right)\\
&=\frac{\sqrt{2}G_{F}}{4\pi R_{\nu}^{2}}\left|
\frac{L_{\nu_{e}}}{\average{E_{\nu_{e}}}}-\frac{L_{\bar{\nu}_{e}}}{\average{E_{\bar{\nu}_{e}}}}
\right|
\left(
1-\sqrt{1-\left(
\frac{R_{\nu}}{r}
\right)^{2}}
\right)^{2},
\end{split}
\end{equation}
where $L_{\nu_{i}}$ and $\average{E_{\nu_{i}}}$ are luminosity and mean energy of $\nu_{i} (\nu_{i}=\nu_{e},\bar{\nu}_{e})$ on the surface of the neutrino sphere. This strength is almost proportional to $r^{-4}$ for large radius ($r>>R_{\nu}$). The small asymmetry between $\nu_{e}$ number flux and that of $\bar{\nu}_{e}$ on the surface of the neutrino sphere would be favorable for the neutrino-antineutrino oscillations because of the small value of $\zeta$.

In outer layers of the supernova material ($r\geq\mathcal{O}(10^{3})$ km), neutrino-neutrino interactions become small and the matter potential is comparable to the vacuum Hamiltonian, which induces the RSF conversions \cite{Ando:2002sk,Ando:2003pj,Yoshida:2009ec}. The contribution from the vacuum potential is no longer negligible in such outer region. Therefore, the neutrino-antineutrino oscillations may be negligible if the strength of vacuum Hamiltonian is larger than that of magnetic field potential. The necessary condition for the equilibration of neutrino-antineutrino oscillations in CCSNe can be summarized as follows:
\begin{equation}
\label{eq:condition eta}
\begin{split}
&\Omega_{\mathrm{mag}}\geq\eta,\\
&\eta=\mathrm{max}\left\{
\lambda,\zeta, \omega
\right\},
\end{split}
\end{equation}
where $\omega=|\Delta m^{2}_{32}|/2E$ is the atmospheric vacuum frequency which characterizes the strength of vacuum Hamiltonian. We impose a typical mean energy of supernova neutrinos: $E=10$ MeV on $\omega$. 

\begin{figure}[t]
\includegraphics[width=0.95\linewidth]{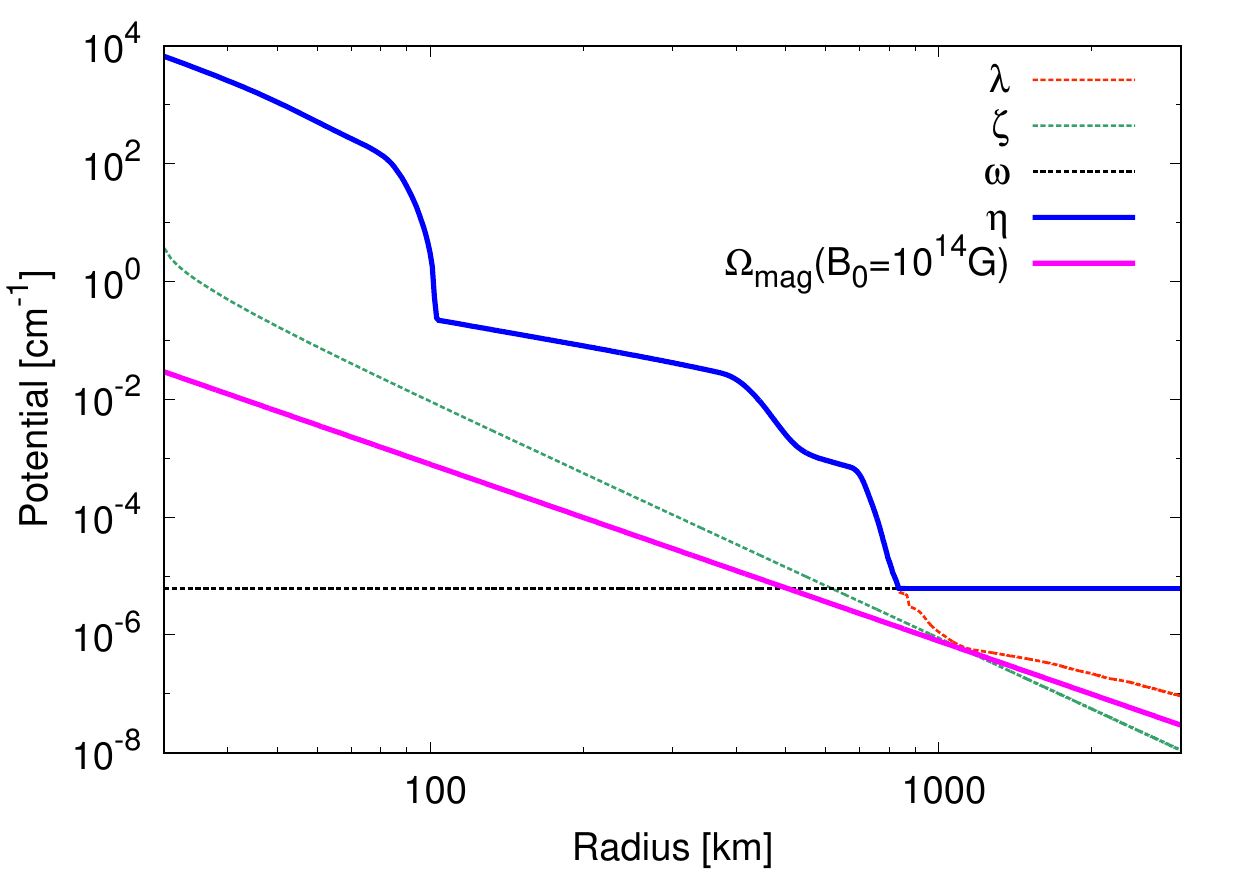}\\
\includegraphics[width=0.95\linewidth]{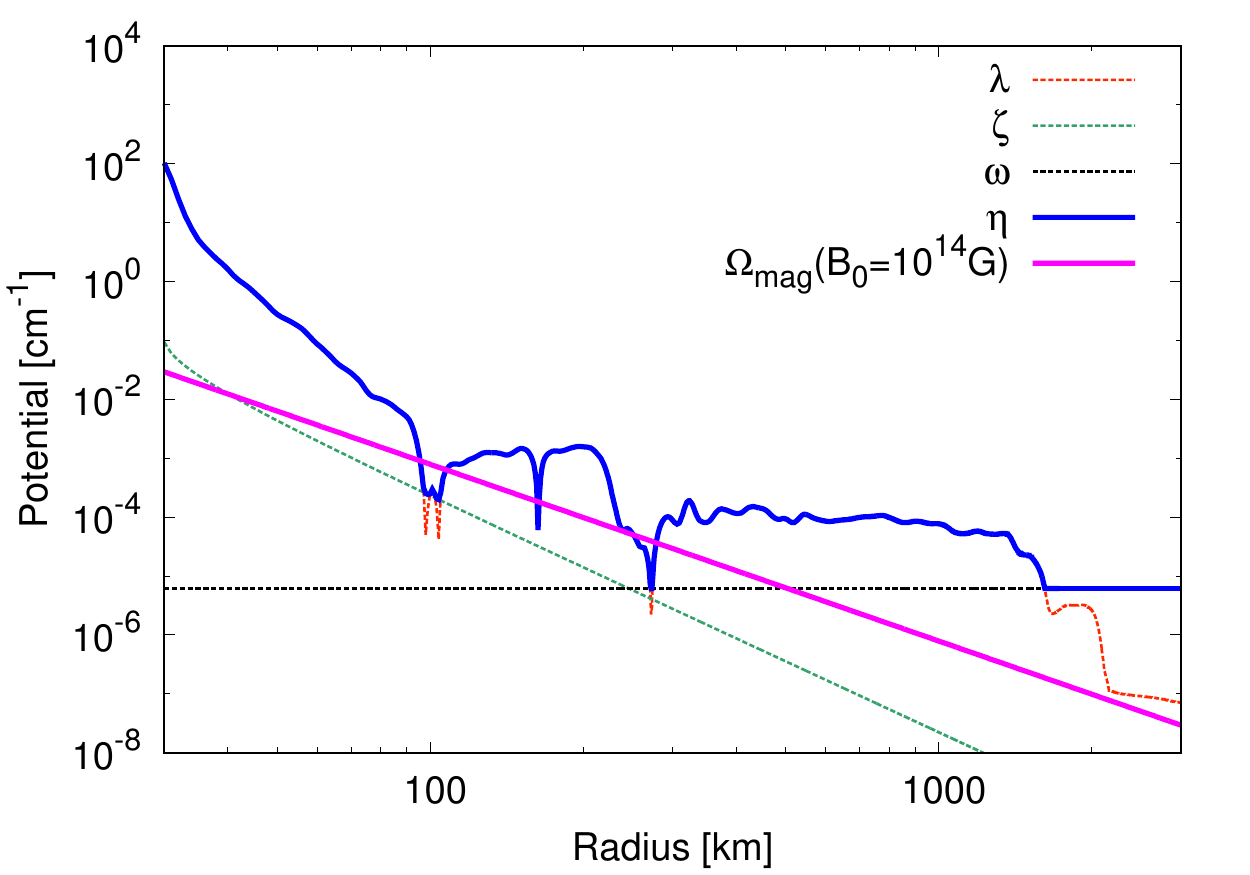}
\caption{%%
%aaaaa
The top (bottom) panel shows strengths of neutrino potentials at $31$ ($331$) ms postbounce, respectively. The potential of the magnetic field (magenta solid line) is given by assuming $B_{0}=10^{14}$G and $\mu_{\nu}=10^{-12}\mu_{B}$ in Eq.(\ref{eq:magnetic field radial profile}). The necessary condition of neutrino-antineutrino oscillations in Eq.(\ref{eq:condition eta}) is realized when $\eta$ (blue solid line) becomes smaller than $\Omega_{\mathrm{mag}}$(magenta solid line).
}
\label{fig:potential CCSN}
\end{figure}

\subsection{
Comparison of potentials in a CCSN model}\label{sec:Comparison of potentials in a CCSN model}

The strengths of neutrino potentials in different explosion phases are shown in Fig.\ref{fig:potential CCSN}. The top panel shows the case of the early explosion phase at $31$ ms postbounce. The strength of the matter potential $\lambda$ is the largest near the surface of the PNS, so that $\eta=\lambda$ is satisfied and the blue solid line completely corresponds to the red dot line in the top panel of Fig.\ref{fig:potential CCSN}. The decrease of $\lambda$ around $100$ km corresponds to the reduction of the baryon density outside the shock wave. There is a boundary between the iron core and Si layer around $400$ km. The matter potential $\lambda\propto |2Y_{e}-1|$ decreases rapidly outside the Si layers because of $Y_{e}\sim0.5$. The matter potential becomes smaller in outer region and finally comparable with the vacuum potential. In outer region ($r>840$ km), the vacuum potential becomes dominant in neutrino Hamiltonian and $\eta$ corresponds to $\omega$. A magenta solid line in the top panel of Fig.\ref{fig:potential CCSN} represents a radial profile of $\Omega_{\mathrm{mag}}=\mu_{\nu}B_{T}$ assuming $B_{0}=10^{14}$ G in Eq.(\ref{eq:magnetic field radial profile}). The necessary condition of the neutrino-antineutrino oscillations is satisfied where the value of $\Omega_{\mathrm{mag}}$ (magenta solid line) is larger than that of $\eta$ (blue solid line). In the top panel of Fig.\ref{fig:potential CCSN}, there is no crossing of these two solid lines, so that Eq.(\ref{eq:condition eta}) is not realized at $31$ ms postbounce in the case of $B_{0}=10^{14}$G. The dense matter profile in the early explosion phase raises up the value of $\eta$ near the PNS. The potential $\zeta$ is  not a dominant term in the early explosion phase of standard CCSNe having iron cores in the progenitors, but we can not ignore the contribution from neutrino-neutrino interactions in the case of an electron capture supernova (ECSN) as shown in Appendix \ref{sec:appendix ECSN potential comparison}. The line of $\Omega_{\mathrm{mag}}$ is shifted upwards by increasing value of $B_{0}$. The top panel of Fig.\ref{fig:potential CCSN} indicates that a stronger magnetic field on the surface of neutrino sphere ($B_{0}>10^{14}$ G) is required for the neutrino-antineutrino oscillations in the early explosion phase.

The radial profiles of potentials at $331$ ms postbounce are shown in the bottom panel of Fig.\ref{fig:potential CCSN}. The values of $\lambda$ and $\zeta$ in the bottom panel of Fig.\ref{fig:potential CCSN} are smaller than those in the top panel of Fig.\ref{fig:potential CCSN} because the baryon density and neutrino fluxes near the PNS decrease as the explosion time has passed. The small values of these potentials enable the crossing of $\eta$ (blue solid line) and $\Omega_{\mathrm{mag}}$ (magenta solid line) in the bottom panel of Fig.\ref{fig:potential CCSN}. The strong magnetic field on the surface of neutrino sphere ($B_{0}>10^{14}$ G) is enough to fulfill Eq.(\ref{eq:condition eta}) at $331$ ms postbounce. There are several peaks in $\lambda$ (red dot line) which correspond to regions of $Y_{e}\sim0.5$. The value of $Y_{e}$ in the supernova material is changed by a heating of the shock wave propagation. Neutrino-neutrino interactions become prominent in the region of $Y_{e}\sim0.5$, which prevents the reduction of $\eta$.

\begin{figure}[t]
\includegraphics[width=0.95\linewidth]{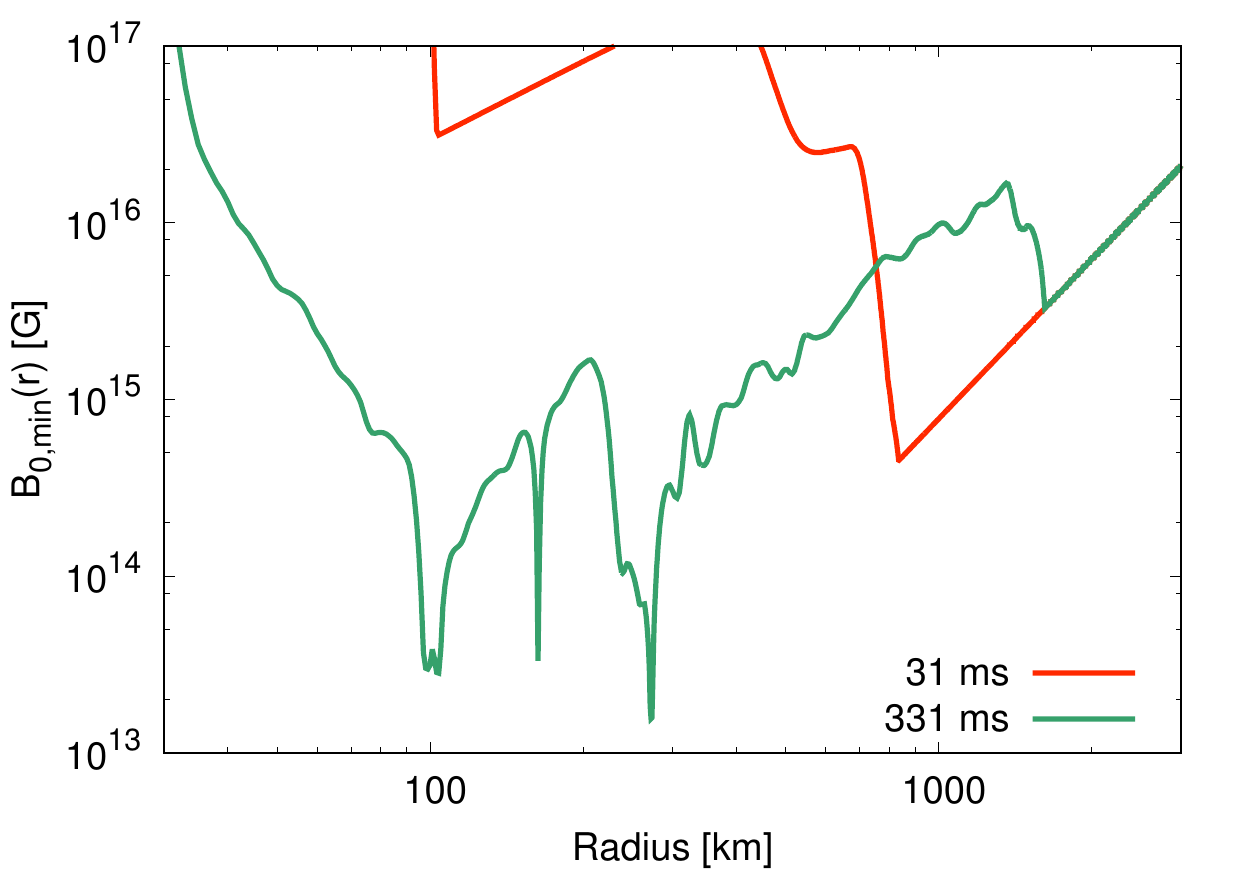}\\
\caption{%%
%aaaaa
Radial profiles of $B_{0,\mathrm{min}}(r)$ at $31$ ms and $331$ ms postbounce. $B_{0,\mathrm{min}}(r)$ is defined in Eq.(\ref{eq:minimum initial magnetic field at r}). The magnetic field at $r=R_{\nu}$ ($B_{0}$) should be larger than $4.52\times10^{14}(1.56\times10^{13})$G in order to satisfy Eq.(\ref{eq:condition eta}) at
$31$($331$) ms, respectively.}
\label{fig:B_min radial profile CCSN}
\end{figure}

In order to satisfy Eq.(\ref{eq:condition eta}) at a radius $r$, the magnetic field on the surface of neutrino sphere $B_{0}$ should be larger than $B_{0,\mathrm{min}}(r)$ written as
\begin{equation}
\label{eq:minimum initial magnetic field at r}
B_{0,\mathrm{min}}(r)=\frac{r^{3}}{\mu_{\nu}R^{3}_{\nu}}\eta.
\end{equation}
The radial profiles of $B_{0,\mathrm{min}}(r)$ at the early and the later explosion phases are shown in Fig.\ref{fig:B_min radial profile CCSN}. As shown in the case of $33$ ms (red solid line), the minimum value of $B_{0,\mathrm{min}}(r)$ is given by $4.52\times10^{14}$G at $r=836$ km. Therefore, $B_{0}$ should be larger than $4.52\times10^{14}$G to fulfill Eq.(\ref{eq:condition eta}) in the early explosion phase. On the other hand, in the case of $331$ ms (green solid line), small baryon density near the PNS reduces the value of $\eta$, which results in smaller value of $B_{0,\mathrm{min}}(r)$ than that in the early phase. Several peaks of $\lambda$ as shown in the bottom panel of Fig.\ref{fig:potential CCSN} induce small values of $B_{0,\mathrm{min}}(r)$ in Fig.\ref{fig:B_min radial profile CCSN}. The necessary condition of Eq.(\ref{eq:condition eta}) is satisfied in the later explosion phase if $B_{0}$ is larger than $1.56\times10^{13}$ G. The required magnetic field for the neutrino-antineutrino oscillations becomes smaller in the later explosion phase, so that the later explosion phase would be favorable to observe the neutrino-antineutrino oscillations. Such feature would be general and independent of progenitor models. The case of ECSN is shown in Appendix \ref{sec:appendix ECSN potential comparison}.\\

\section{Summary and Discussion} 
\label{sec:Summary and Discussion}

We study neutrino-antineutrino oscillations caused by a strong magnetic field in dense matter in the case of Majorana neutrinos. Numerical simulations of such neutrino-antineutrino oscillations are carried out by changing values of baryon densities and electron fractions. We reveal that the neutrino-antineutrino oscillations are sensitive to strengths of matter potentials such as $|\lambda_{e}-\lambda_{n}|$ and $\lambda_{n}$. The flavor equilibration occurs when the strength of magnetic field potential is comparable with that of a matter potential: $\Omega_{\mathrm{mag}}\sim\lambda_{n}$. Equilibrium states of such three flavor neutrino-antineutrino oscillations are consistent with numerical results in Ref.\cite{Abbar:2020ggq}. On the other hand, in the case of more dense matter ($|\lambda_{e}-\lambda_{n}|, \lambda_{n} >> \Omega_{\mathrm{mag}}$), any flavor conversions are suppressed because correlations between neutrinos and antineutrinos fail to grow up in large matter potentials. The values of matter potentials depend on both the baryon density and electron fractions inside material. We find that the flavor equilibration also appears around $Y_{e}\sim0.5$ irrespective of large baryon densities. One of the neutrino flavors is decoupled from flavor conversions around $Y_{e}\sim0.5$ in dense matter. The values of equilibrium states of $\average{\rho_{ee}}$ and $\average{\bar{\rho}_{ee}}$ are different from those in three flavor conversions in neutron (proton)-rich matter.

The mechanism of the neutrino-antineutrino oscillations in strong magnetic fields, which we focus here is different from that of RSF. In our simulations, the matter potential is much higher than the vacuum Hamiltonian, so that there is no resonance like RSF. The equilibration of neutrino-antineutrino oscillations in our simulations is independent of neutrino mass hierarchy and neutrino energy. Such neutrino-antineutrino oscillations are sensitive to the strength of neutrino-neutrino interactions \cite{Abbar:2020ggq}. As shown in Appendix \ref{sec:appendix analytical explanation}, however, the flavor equilibration of neutrino-antineutrino oscillations is possible even in the absence of neutrino-neutrino interactions. The origin of the neutrino-antineutrino oscillations is the coupling of matter potentials with the magnetic field potential.

We discuss the possibility to occur the equilibration of neutrtino-antineutrino oscillations in a $11.2M_{\odot}$ CCSN model. Here, we utilize a necessary condition for the neutrino-antineutrino oscillations in CCSNe. In order to satisfy such condition, the magnetic field on the surface of the PNS should be larger than $10^{13}$G in the case of $\mu_{\nu}=10^{-12}\mu_{B}$ which is the same order as a tight upper limit of neutrino magnetic moment \cite{Arceo-Diaz:2015pva}. The strength of matter potential becomes smaller near the PNS as the explosion phase has passed, so that the required magnetic field for the neutrino-antineutrino oscillations becomes smaller in the later explosion phase. Such trend is also confirmed in ECSN irrespective of different matter profiles from an iron core of the standard CCSNe (see Appendix \ref{sec:appendix ECSN potential comparison}).

Our result indicates that the typical magnetic field of pulsars ($\sim10^{12}$G) is not enough to induce the flavor equilibrium of neutrino-antineutrino oscillations near the PNS. On the other hand, the neutrino-antineutrino oscillations are possible in the typical magnetic field of magnetars ($\sim10^{14}$G) \cite{Enoto:2019}. The strength of the magnetic field potential $\Omega_{\mathrm{mag}}$ is proportional to neutrino magnetic moments and the transverse magnetic field, so that a stronger magnetic field can induce the neutrino-antineutrino oscillations even in $\mu_{\nu}<10^{-12}\mu_{B}$. A supernova explosion scenario leaving an magnetar at the center potentially updates the current upper limit on $\mu_{\nu}$. Further quantitative studies discussing the magnetic field effect on neutrino detection will be required to identify possibilities to withdraw properties of neutrinos from supernova neutrinos. Here, we focus on specific supernova progenitor models, but the necessary condition of the neutrino-antineutrino oscillations proposed in this work can be applied in more general explosive phenomena such as neutron-star mergers and gamma-ray bursts. \\

\begin{acknowledgments}
This study was supported in part by JSPS/MEXT KAKENHI Grant Numbers JP19J13632, % Sasaki
JP18H01212, % Yokoi Kinban B (Takiwaki Co-PI)
JP17H06364, %Shingakujutu GWGEN C01,  (Takiwaki Co-PI).
JP21H01088.
This work is also supported by the NINS program for cross-disciplinary
study (Grant Numbers 01321802 and 01311904) on Turbulence, Transport,
and Heating Dynamics in Laboratory and Solar/Astrophysical Plasmas:
"SoLaBo-X”.
Numerical computations were carried out on PC cluster and Cray XC50 at the Center for Computational Astrophysics,
National Astronomical Observatory of Japan.
This research was also supported by MEXT as “Program for Promoting 
researches on the Supercomputer Fugaku” (Toward a unified view of 
the universe: from large scale structures to planets) and JICFuS.

\end{acknowledgments}

\appendix

\section{The matter suppression in neutrino-antineutrino oscillations}
\label{sec:appendix analytical explanation}

We discuss the mechanism of the flavor equilibration of neutrino-antineutrino oscillations in strong magnetic fields, which is confirmed in Figs.\ref{fig:matter_dep} and \ref{fig:ye_dep}. Here, we assume that the neutrino-antineutrino oscillations are dominant and ordinary flavor conversions without magnetic fields are negligible. Furthermore, we consider flavor conversions in dense matter where a vacuum potential $\Omega(E)$ and a potential of neutrino-neutrino interactions is negligible. Such conditions are written as
\begin{equation}
\label{eq:communicate matrix}
\begin{split}
\left[
H_{\nu},\rho_{\theta}
\right]&\sim 0,\\
\left[
H_{\bar{\nu}},\bar{\rho}_{\theta}\right]&\sim 0,\\
H_{\nu}\sim H_{\bar{\nu}}&\sim -V_{\mathrm{matter}},\\
\end{split}
\end{equation}
where $H_{\nu}$ ($H_{\bar{\nu}}$) is the Hamiltonian of neutrinos (antineutrinos), respectively, without the magnetic field potential. We focus on flavor conversions of neutrinos whose emission angle is $\theta=0$. The $\theta=0$ is imposed on Eq.(\ref{eq:Liouville-von Neumann}) and the index $\theta$ in a neutrino density matrix is dropped hereafter. We investigate flavor conversions in $e-x-y$ basis \cite{Dasgupta:2007ws} instead of the flavor $e-\mu-\tau$ basis. The matter potential is invariant under the rotation from flavor basis to $e-x-y$ basis:
\begin{equation}
\begin{split}
V^{\prime}_{\mathrm{matter}}&=R^{T}(\theta_{23})V_{\mathrm{matter}}R(\theta_{23})=V_{\mathrm{matter}},
\end{split}
\end{equation}
\begin{equation}
R(\theta_{23})=\left(
\begin{array}{c c c}
1&0&0\\
0&\cos\theta_{23}&\sin\theta_{23}\\
0&-\sin\theta_{23}&\cos\theta_{23}
\end{array}\right).
\end{equation}
On the other hands, the magnetic field potential does not commute with $R(\theta_{23})$, so that the matrix components are transformed by the rotation from the flavor basis to the $e-x-y$ basis:
\begin{equation}
\label{eq:magnetic potential e-x-y}
\begin{split}
V_{\mathrm{mag}}^{\prime}&=R^{T}(\theta_{23})V_{\mathrm{mag}}R(\theta_{23})\\
&=\Omega_{\mathrm{mag}}\left(
\begin{array}{c c c}
0&0&1\\
0&0&1\\
-1&-1&0
\end{array}\right),
\end{split}
\end{equation}
where $\Omega_{\mathrm{mag}}=\mu_{\nu}B_{T}$. The evolutions of neutrino density matrices in Eq.(\ref{eq:Liouville-von Neumann}) are decomposed by
\begin{equation}
\partial_{r}\rho=-i\left[
H_{\nu},\rho
\right]-i\left(
V_{\mathrm{mag}}X^{\dagger}+XV_{\mathrm{mag}}
\right),
\end{equation}
\begin{equation}
\partial_{r}\bar{\rho}=-i\left[
H_{\bar{\nu}},\bar{\rho}
\right]+i\left(
V_{\mathrm{mag}}X+X^{\dagger}V_{\mathrm{mag}}
\right),
\end{equation}
\begin{equation}
\label{eq:evolution of X}
\partial_{r}X=-i\left(
H_{\nu}X-XH_{\bar{\nu}}+V_{\mathrm{mag}}\bar{\rho}-\rho V_{\mathrm{mag}}
\right).
\end{equation}
From Eqs.(\ref{eq:communicate matrix})-(\ref{eq:evolution of X}), evolutions of diagonal components of neutrino density matrices are described by
\begin{equation}
\label{eq:evolution ee e-x-y}
\partial_{r}\rho_{ee}\sim-2\Omega_{\mathrm{mag}}X_{ey,i},
\end{equation}
\begin{equation}
\label{eq:evolution xx e-x-y}
\partial_{r}\rho_{xx}\sim-2\Omega_{\mathrm{mag}}X_{xy,i},
\end{equation}
\begin{equation}
\label{eq:evolution yy e-x-y}
\partial_{r}\rho_{yy}\sim2\Omega_{\mathrm{mag}}\left(
X_{ye,i}+X_{yx,i}
\right),
\end{equation}
\begin{equation}
\label{eq:evolution ee e-x-y anti}
\partial_{r}\bar{\rho}_{ee}\sim-2\Omega_{\mathrm{mag}}X_{ye,i},
\end{equation}
\begin{equation}
\label{eq:evolution xx e-x-y anti}
\partial_{r}\bar{\rho}_{xx}\sim-2\Omega_{\mathrm{mag}}X_{yx,i},
\end{equation}
\begin{equation}
\label{eq:evolution yy e-x-y anti}
\partial_{r}\bar{\rho}_{yy}\sim2\Omega_{\mathrm{mag}}\left(
X_{ey,i}+X_{xy,i}
\right),
\end{equation}
where $X_{\alpha\beta,i} (\alpha,\beta=e,x,y)$ represents the imaginary parts of $X_{\alpha\beta}$. From these equation of motions, we can find conservation laws:
\begin{equation}
\label{eq:conservation laws}
\begin{split}
\rho_{ee}+\rho_{xx}+\bar{\rho}_{yy}&=\mathrm{const}.,\\
\bar{\rho}_{ee}+\bar{\rho}_{xx}+\rho_{yy}&=\mathrm{const}.,
\end{split}
\end{equation}
which result in a decoupling of the $\nu_{e}-\nu_{x}-\bar{\nu}_{y}$ sector and the $\bar{\nu}_{e}-\bar{\nu}_{x}-\nu_{y}$ sector during the neutrino-antineutrino oscillations. The above conservation laws are actually confirmed in our numerical simulations. Here, we only focus on flavor conversions among $\nu_{e},\nu_{x}$ and $\bar{\nu}_{y}$. Almost the same discussion is possible in the $\bar{\nu}_{e}-\bar{\nu}_{x}-\nu_{y}$ sector because of the decoupling in Eq.(\ref{eq:conservation laws}). We require evolution of $X_{ey,i}$ and $X_{xy,i}$ in order to close Eqs.(\ref{eq:evolution ee e-x-y}), (\ref{eq:evolution xx e-x-y}) and (\ref{eq:evolution yy e-x-y anti}). Non-diagonal components such as $\rho_{ex},\rho_{ey},\bar{\rho}_{ex}$, and $\bar{\rho}_{ey}$ would be ignored under the assumptions of Eq.(\ref{eq:communicate matrix}). Therefore, equation of motions of $X_{ey}$ and $X_{xy}$ are given by
\begin{equation}
\partial_{r}X_{ey,r}\sim\left(
\lambda_{e}-\lambda_{n}
\right)X_{ey,i},
\end{equation}
\begin{equation}
\label{eq:evolution X eyi}
\partial_{r}X_{ey,i}\sim-\left(
\lambda_{e}-\lambda_{n}
\right)X_{ey,r}-\Omega_{\mathrm{mag}}\left(
\bar{\rho}_{yy}-\rho_{ee}
\right),
\end{equation}
\begin{equation}
\partial_{r}X_{xy,r}\sim-\lambda_{n}X_{xy,i},
\end{equation}
\begin{equation}
\label{eq:evolution X xyi}
\partial_{r}X_{xy,i}\sim\lambda_{n}X_{xy,r}-\Omega_{\mathrm{mag}}\left(
\bar{\rho}_{yy}-\rho_{xx}
\right),
\end{equation}
where $X_{ey,r}(X_{xy,r})$ are the real parts of $X_{ey}(X_{xy})$, respectively. Numerical results of our simulations can be analyzed by comparing three different frequencies such as $|\lambda_{e}-\lambda_{n}|$, $\lambda_{n}$, and $\Omega_{\mathrm{mag}}$ in the above differential equations. \\

\subsection{Case of $|\lambda_{e}-\lambda_{n}|$, $\lambda_{n}>>$ $\Omega_{\mathrm{mag}}$}
\label{sec:complete matter suppression}

In this case, the second term on the right hand side of Eq.(\ref{eq:evolution X xyi}) is almost negligible, so that
\begin{equation}
\partial_{r}^{2}X_{xy,i}\sim-\lambda_{n}^{2}X_{xy,i},
\end{equation}
is obtained. There is no correlation between neutrinos and antineutrinos at the beginning of the calculation, so that $X_{xy,i}$ should be proportional to $\sin(\lambda_{n}r)$. The coefficient of $X_{xy,i}$ can be derived by substituting $X_{xy,i}$ for Eq.(\ref{eq:evolution X xyi}) and consider the case of $r=0$. However, the right hand side of Eq.(\ref{eq:evolution X xyi}) is zero at $r=0$. Therefore, $X_{xy,i}$ is also zero irrespective of the radius, which results in the decoupling of $\nu_{x}$ during the neutrino-antineutrino oscillations. It is also assumed that the 
$|\lambda_{e}-\lambda_{n}|$ is much larger than $\Omega_{\mathrm{mag}}$. This means the value of $|2Y_{e}-1|$ is finite. The derivation of $X_{ey,i}$ is carried out in analogy with $X_{xy,i}$. The value of $X_{ey,i}$ is almost zero because the second term on the right hand side of Eq.(\ref{eq:evolution X eyi}) is negligible. Therefore, matter suppression is dominant and any flavor conversion does not appear. This analytical model can explain the strong matter suppression in our simulations (e.g. $\lambda_{e}=10^{2}\Omega_{\mathrm{mag}}$(dark-orange solid lines) in Fig.\ref{fig:matter_dep}).\\
 
\begin{figure}[t]
\includegraphics[width=0.95\linewidth]{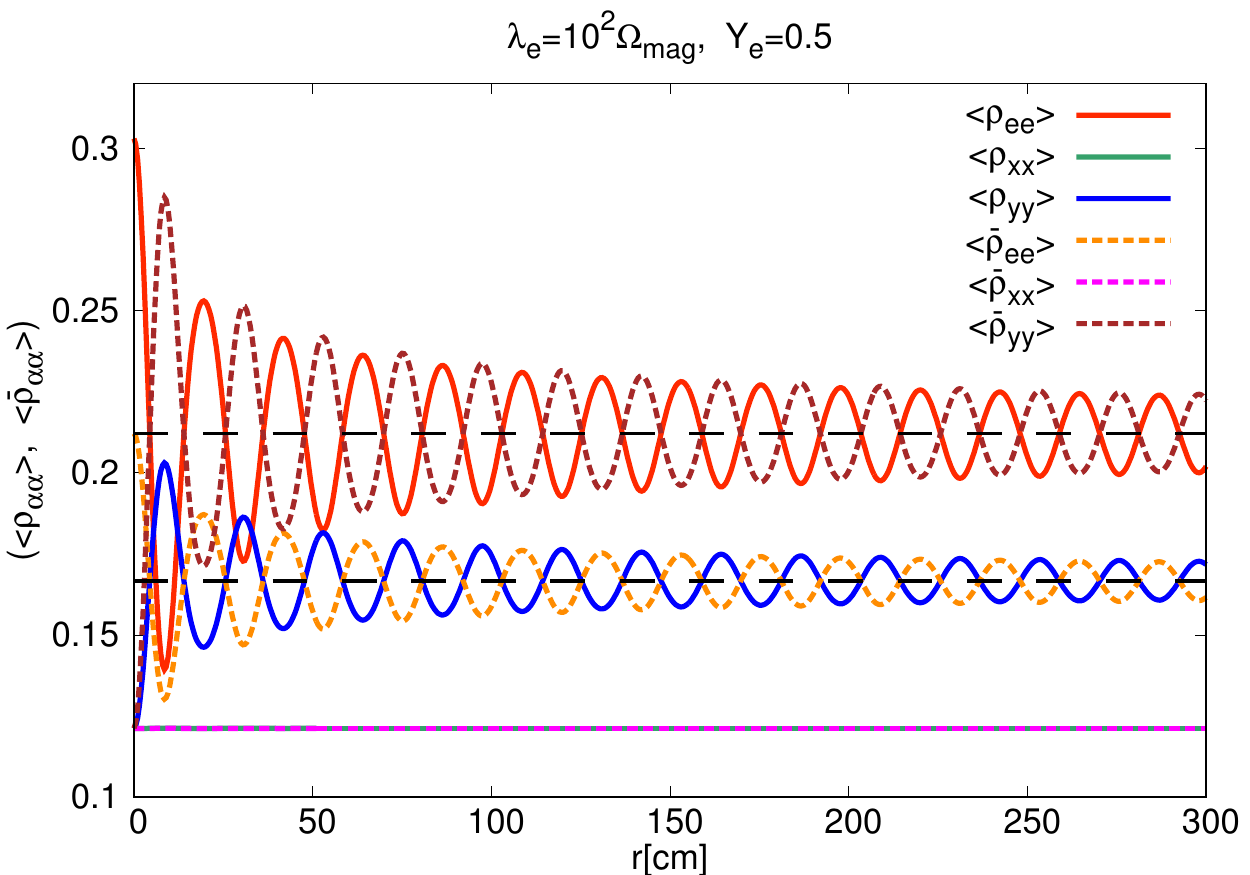}
\caption{%%
The evolution of angle averaged diagonal components of neutrino density matrices. We employ the same parameter set as that in the case of $Y_{e}=0.5$ in Fig.\ref{fig:ye_dep}. The constant values of the black dash lines correspond to equilibrium values of the two flavor neutrino-antineutrino oscillations such as $(n_{\nu_{e}}+n_{\nu_{x}})/2n_{n_{\nu}}=7/33$ and $(n_{\bar{\nu}_{e}}+n_{\nu_{x}})/2n_{n_{\nu}}=1/6$.
}
\label{fig:density matrix 2flavor}
\end{figure}

\subsection{Case of $\lambda_{e}=\lambda_{n}(Y_{e}=0.5)$, $\lambda_{n}>> \Omega_{\mathrm{mag}}$}
 \label{sec:one flavor decoupling in three flavor oscillations}
 
The $X_{xy,i}$ is negligible and $\nu_{x}$ decouples from the flavor conversions as shown in Sec.\ref{sec:complete matter suppression}. The difference from the previous case is $|\lambda_{e}-\lambda_{n}|=0$ where the electron fraction $Y_{e}$ is equal to $0.5$. The finite $X_{ey,i}$ induces two flavor oscillations between $\nu_{e}$ and $\bar{\nu}_{y}$ irrespective of a large matter potential: $\lambda_{n}>>\Omega_{\mathrm{mag}}$. The second derivative of $X_{ey,i}$ is described by
\begin{equation}
\label{eq:second derivative X eyi 2 flavor}
\partial_{r}^{2}X_{ey,i}\sim-4\Omega_{\mathrm{mag}}^{2}X_{ey,i},
\end{equation}
where ($\bar{\rho}_{yy}-\rho_{ee}$) in Eq.(\ref{eq:evolution X eyi}) is eliminated by multiplying the derivative $\partial_{r}$ and using Eqs.(\ref{eq:evolution yy e-x-y anti}) and (\ref{eq:evolution ee e-x-y}). The correlation $X_{ey,i}$ is derived by solving Eq.(\ref{eq:second derivative X eyi 2 flavor}) and imposing initial condition of neutrino density matrices on Eq.(\ref{eq:evolution X eyi}). In our simulations, the initial condition of the neutrino density matrix at $r=0$ is given by $\rho_{ee}=n_{\nu_{e}}/n_{\nu}$, $\rho_{xx}=n_{\nu_{x}}/n_{\nu}$ and $\bar{\rho}_{yy}=n_{\nu_{x}}/n_{\nu}$ where $n_{\nu}=n_{\nu_{e}}+n_{\bar{\nu}_{e}}+4n_{\nu_{x}}$. Then, by solving Eqs.(\ref{eq:evolution ee e-x-y}) and (\ref{eq:evolution yy e-x-y anti}), 
we obtain
\begin{equation}
\label{eq:2flavor ye=0.5}
\begin{split}
\rho_{ee}&=\frac{n_{\nu_{e}}}{n_{\nu}}+\frac{n_{\nu_{x}}-n_{\nu_{e}}}{2n_{\nu}}\left(
1-\cos2\Omega_{\mathrm{mag}}r
\right),\\
\rho_{xx}&=\frac{n_{\nu_{x}}}{n_{\nu}}=\mathrm{const}.,\\
\bar{\rho}_{yy}&=\frac{n_{\nu_{x}}}{n_{\nu}}-\frac{n_{\nu_{x}}-n_{\nu_{e}}}{2n_{\nu}}\left(
1-\cos2\Omega_{\mathrm{mag}}r
\right).
\end{split}
\end{equation}
The $\rho_{xx}$ is constant and the $\rho_{ee}$ oscillates around $(n_{\nu_{e}}+n_{\nu_{x}})/2n_{n_{\nu}}$ in this analytical model. Our derivation can be easily applied to the $\bar{\nu}_{e}-\bar{\nu}_{x}-\nu_{y}$ sector by solving evolutions of $X_{ye,i}$ and $X_{yx,i}$ instead of $X_{ey,i}$ and $X_{xy,i}$. We can show that flavor conversions between $\bar{\nu}_{e}$ and $\nu_{y}$ occur around an equilibrium value $(n_{\bar{\nu}_{e}}+n_{\nu_{x}})/2n_{n_{\nu}}$ and $\bar{\nu}_{x}$ decouples from flavor conversions. The above analytical discussion reproduces numerical results of Fig.\ref{fig:density matrix 2flavor} qualitatively. Fig.\ref{fig:density matrix 2flavor} represents evolution of angle averaged diagonal components of neutrino density matrices as defined in Eqs.(\ref{eq:angle average rho_e}) and (\ref{eq:angle average rho_eb}) in $e-x-y$ basis. The input parameters used in Fig.\ref{fig:density matrix 2flavor} are the same as those in the case of $Y_{e}=0.5$ in Fig.\ref{fig:ye_dep} (blue solid line). Fig.\ref{fig:density matrix 2flavor} shows that one of the flavors $x$ is decoupled from neutrino oscillations and two flavor neutrino-antineutrino oscillations in $\nu_{e}-\bar{\nu}_{y}$ and $\nu_{y}-\bar{\nu}_{e}$ sectors happen around $(n_{\nu_{e}}+n_{\nu_{x}})/2n_{n_{\nu}}$ and $(n_{\bar{\nu}_{e}}+n_{\nu_{x}})/2n_{n_{\nu}}$, respectively. The decreasing oscillation amplitudes of $\average{\rho_{\alpha\alpha}}$ and  $\average{\bar{\rho}_{\alpha\alpha}}$ $(\alpha=e,y)$ up to $r\sim100$ cm would originate from the angle average on neutrino density matrices.\\

\begin{figure}[t]
\includegraphics[width=0.95\linewidth]{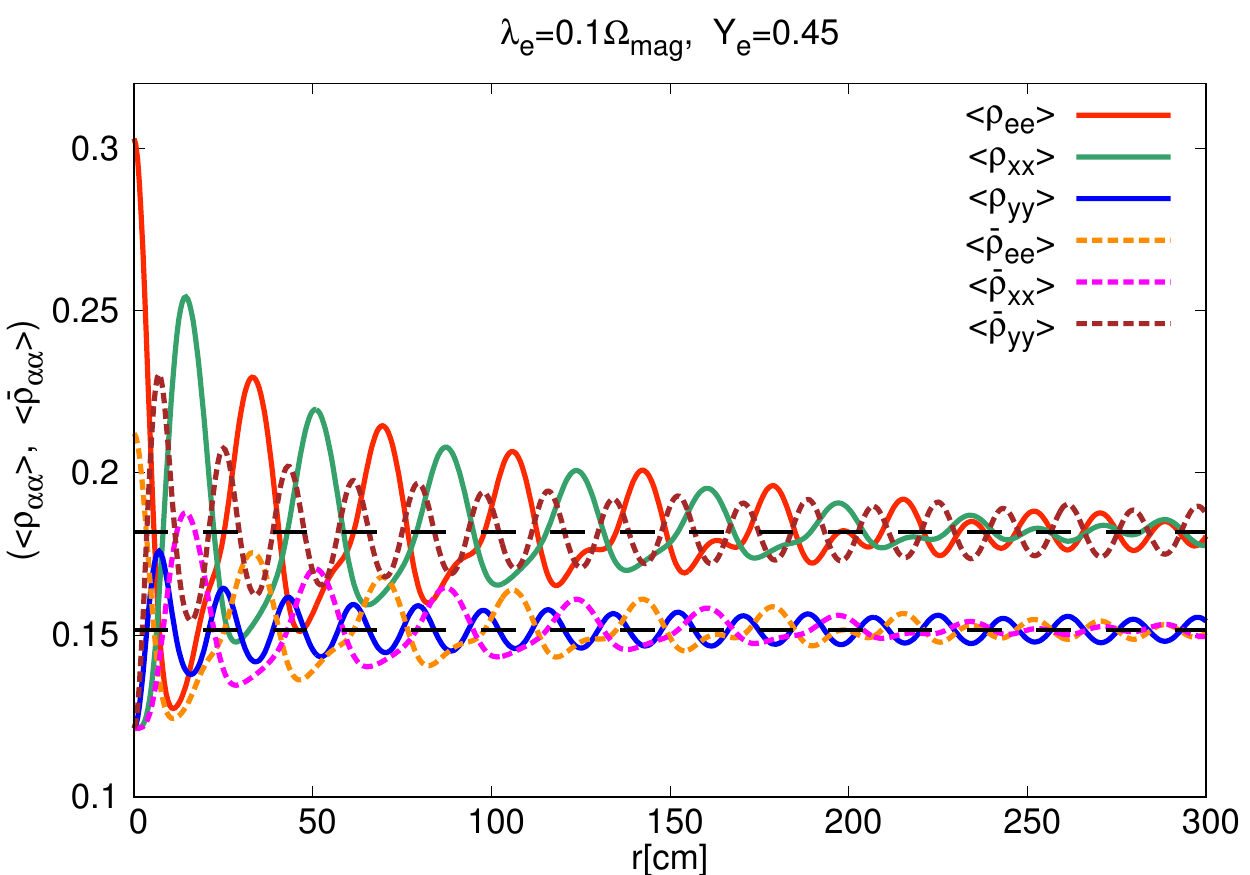}\\
\caption{%%
The evolution of diagonal components like Fig.\ref{fig:density matrix 2flavor}. The input parameters are the same as that in the case of $\lambda_{e}=0.1\Omega_{\mathrm{mag}}$ in Fig.\ref{fig:matter_dep}. The constant black dash lines show equilibrium values of the three flavor neutrino-antineutrino oscillations such as $(n_{\nu_{e}}+2n_{\nu_{x}})/3n_{\nu}=2/11$ and $(n_{\bar{\nu}_{e}}+2n_{\nu_{x}})/3n_{\nu}=5/33$.}
\label{fig:density matrix 3flavor}
\end{figure}

\subsection{Case of $\Omega_{\mathrm{mag}}>>$ $|\lambda_{e}-\lambda_{n}|$, $\lambda_{n}$}
\label{sec:appendix three flavor}

The matter potentials are smaller than the magnetic field potential, so that the contribution from $X_{xy,i}$ is no longer negligible and $\nu_{x}$ joins flavor conversions. In this case, the second derivative of $X_{ey,i}$ and $X_{xy,i}$ are written as
\begin{equation}
\begin{split}
\partial_{r}^{2}X_{ey,i}&\sim-4\Omega_{\mathrm{mag}}^{2}X_{ey,i}-2\Omega_{\mathrm{mag}}^{2}X_{xy,i},\\
\partial_{r}^{2}X_{xy,i}&\sim-2\Omega_{\mathrm{mag}}^{2}X_{ey,i}-4\Omega_{\mathrm{mag}}^{2}X_{xy,i}.
\end{split}
\end{equation}
The above differential equations can be solved analytically and coefficients of two modes: $\sin(\sqrt{2}\Omega_{\mathrm{mag}}r)$ and $\sin(\sqrt{6}\Omega_{\mathrm{mag}}r)$ are determined by imposing the initial condition of neutrino density matrices on Eqs.(\ref{eq:evolution X eyi}) and (\ref{eq:evolution X xyi}). The diagonal terms of neutrino density
matrices are given by
\begin{equation}
\begin{split}
\rho_{ee}=\frac{n_{\nu_{e}}}{n_{\nu}}&+\frac{n_{\nu_{x}}-n_{\nu_{e}}}{6n_{\nu}}\left(
1-\cos\sqrt{6}\Omega_{\mathrm{mag}}r
\right)\\
&+\frac{n_{\nu_{x}}-n_{\nu_{e}}}{2n_{\nu}}\left(
1-\cos\sqrt{2}\Omega_{\mathrm{mag}}r
\right),\\
\\
\rho_{xx}=\frac{n_{\nu_{x}}}{n_{\nu}}&+\frac{n_{\nu_{x}}-n_{\nu_{e}}}{6n_{\nu}}\left(
1-\cos\sqrt{6}\Omega_{\mathrm{mag}}r
\right)\\
&-\frac{n_{\nu_{x}}-n_{\nu_{e}}}{2n_{\nu}}\left(
1-\cos\sqrt{2}\Omega_{\mathrm{mag}}r
\right),\\
\\
\bar{\rho}_{yy}=\frac{n_{\nu_{x}}}{n_{\nu}}&-\frac{n_{\nu_{x}}-n_{\nu_{e}}}{3n_{\nu}}\left(
1-\cos\sqrt{6}\Omega_{\mathrm{mag}}r
\right).
\end{split}
\end{equation}
These diagonal components oscillate around $(n_{\nu_{e}}+2n_{\nu_{x}})/3n_{\nu}$. Such analytical model help understand numerical results of three flavor oscillations as shown in our simulations (e.g. $\lambda_{e}=0.1\Omega_{\mathrm{mag}}$(red solid line) in the top panle of Fig.\ref{fig:matter_dep}). In the $\bar{\nu}_{e}-\bar{\nu}_{x}-\nu_{y}$ sector, three flavor oscillations around $(n_{\bar{\nu}_{e}}+2n_{\nu_{x}})/3n_{\nu}$ are obtained analytically in the same way as the above derivation. Fig.\ref{fig:density matrix 3flavor} shows numerical results of the three flavor neutrino-antineutrino oscillations. There are the three flavor neutrino-antineutrino oscillations in both $\nu_{e}-\nu_{x}-\bar{\nu}_{y}$ and $\bar{\nu}_{e}-\bar{\nu}_{x}-\nu_{y}$ sectors as discussed in our analytical treatment. The equilibrium values of flavor conversions in Fig.\ref{fig:density matrix 3flavor} are different from that in Fig.\ref{fig:density matrix 2flavor} because of the finite couplings of $\nu_{x}$ and $\bar{\nu}_{x}$ with the neutrino-antineutrino oscillations. \\

\section{Neutrino potentials in an electron capture supernova (ECSN)}
\label{sec:appendix ECSN potential comparison}

\begin{figure}[t]
\includegraphics[width=0.95\linewidth]{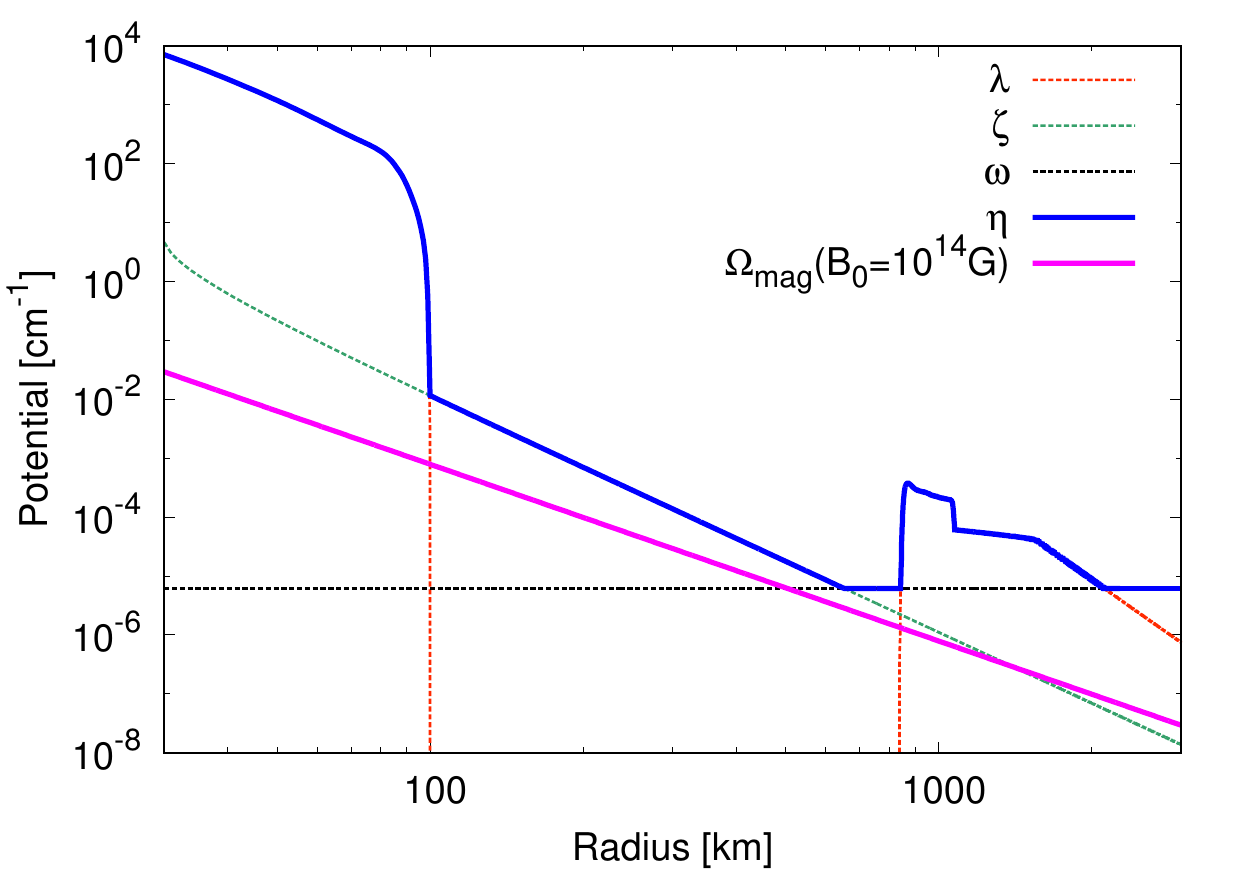}\\
\includegraphics[width=0.95\linewidth]{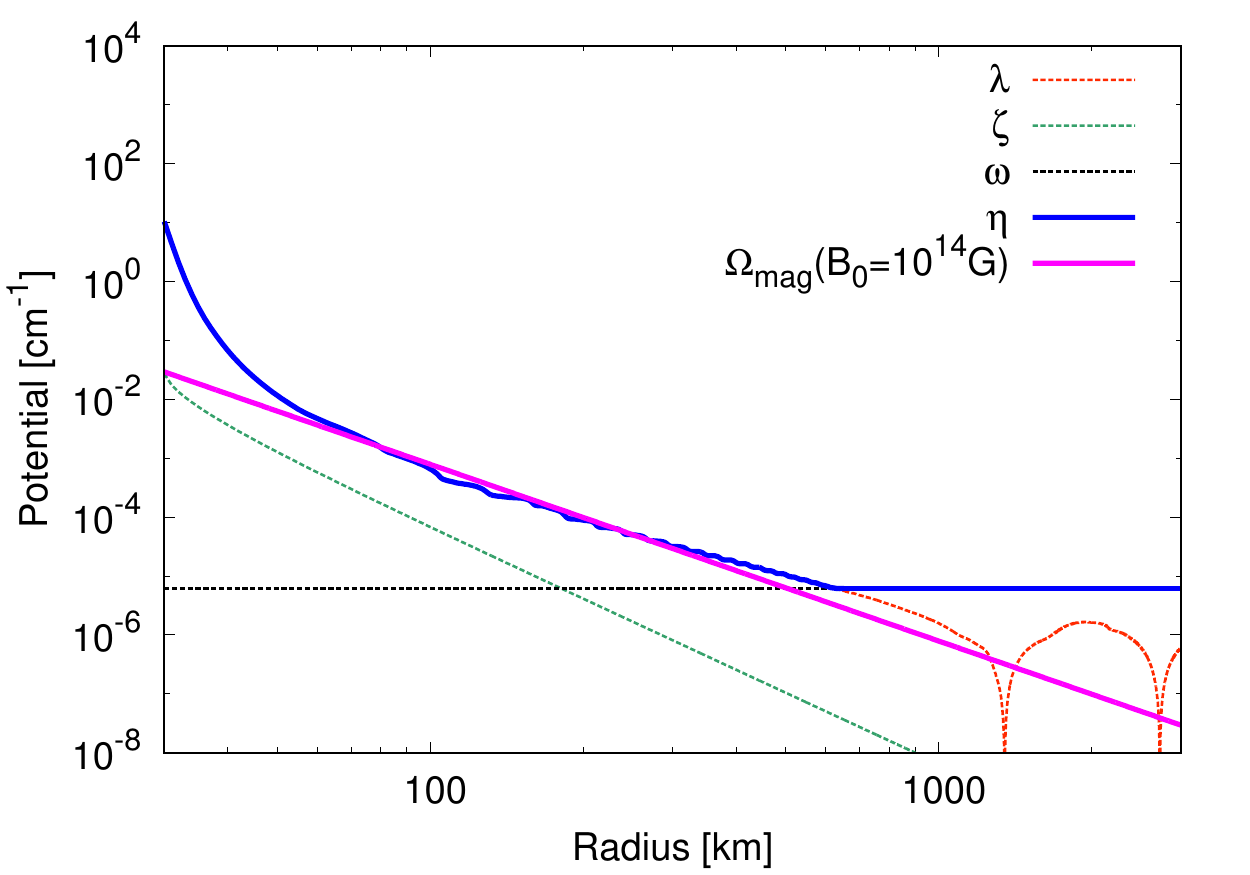}
\caption{%%
%aaaaa
The radial profiles of neutrino potentials in ECSN as in Fig.\ref{fig:potential CCSN}. The top (bottom) panel shows the case at $31$ ($331$) ms postbounce, respectively.
}
\label{fig:potential ECSN}
\end{figure}

\begin{figure}[t]
\includegraphics[width=0.95\linewidth]{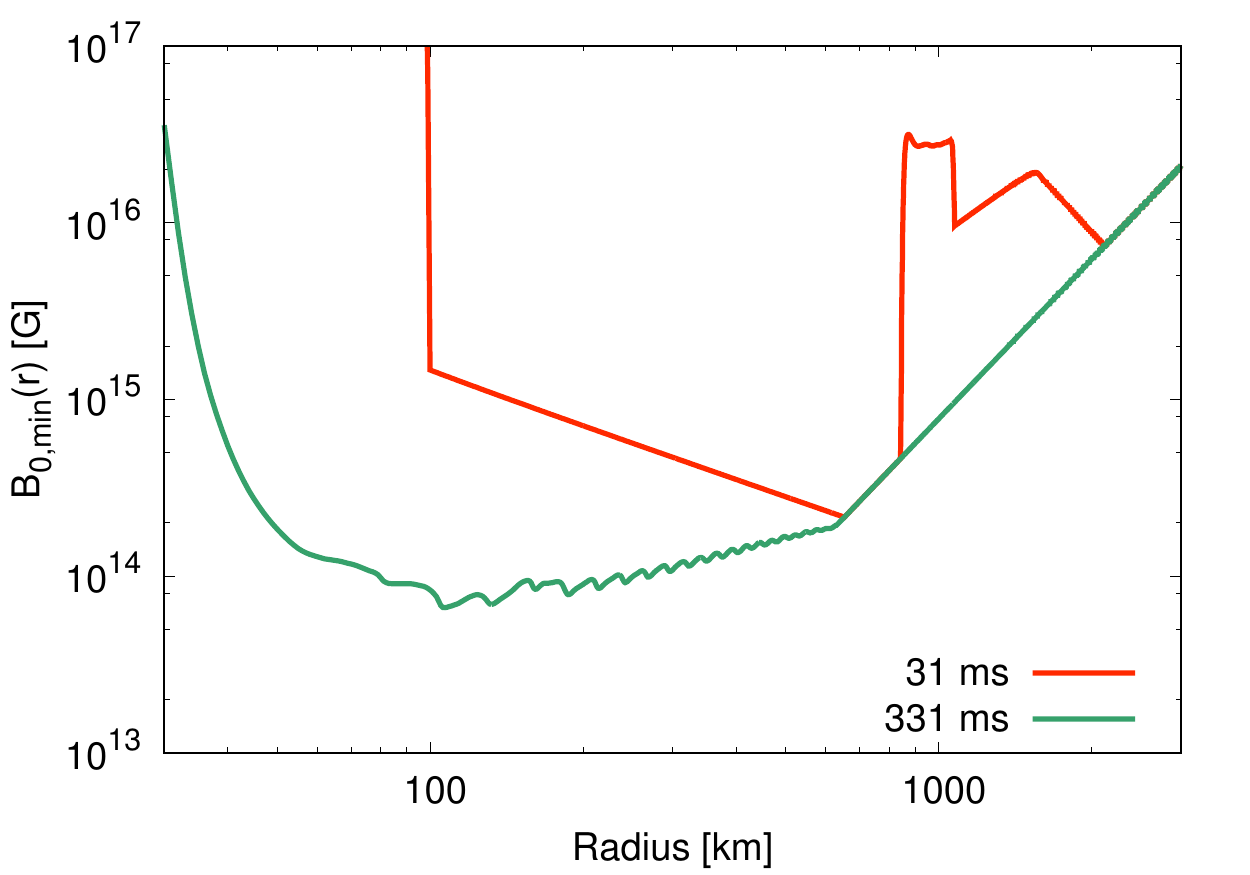}\\
\caption{%%
%aaaaa
Radial profiles of  Eq.(\ref{eq:minimum initial magnetic field at r}) at $31$ ms and $331$ ms postbounce in ECSN.
}
\label{fig:B_min radial profile ECSN}
\end{figure}
This section shows how strong the magnetic field is necessary to induce the flavor equilibration in ECSN,
which has an O-Ne-Mg core in the progenitor. This section is parallel to Section \ref{sec:Comparison of potentials in a CCSN model}, which we discuss the iron core progenitor. We employ the supernova model with an $8.8\,{M_{\odot}}$ progenitor, which is the same to that of Ref.~\cite{Sasaki:2019jny}.

In the early explosion phase at $31$ ms postbounce (the top panel of Fig.\ref{fig:potential ECSN} ), there is a sharp drop of $\lambda$ around $r=100$ km which corresponds to the position of the propagating shock wave. The baryon density decreases significantly outside the shock front. The shock wave has reached a layer of $\alpha$-elements where the electron fraction is $Y_{e}\sim0.5$. The matter potential $\lambda\propto |2Y_{e}-1|$ disappears and neutrino-neutrino interactions become dominant in such region and $\eta$ is equivalent to $\zeta$. The enhancement of the $\eta$ at $r=838$ km would reflect the large electron fraction of $\mathrm{He}+\mathrm{H}$ layers in our progenitor model. The value of $\lambda$ is sensitive to composition of nuclear species in supernova material because of the $Y_{e}$ dependence. Therefore, the radial profile of $\lambda$ as shown in the top panel of Fig.\ref{fig:potential ECSN} would be a unique structure of ECSN which has an O-Mg-Ne core in the progenitor. In outer region ($r>2120$ km), $\eta=\omega$ is satisfied because of the decreasing baryon density and neutrino fluxes. The value of $\eta$ (blue solid line) is always larger than that of $\Omega_{\mathrm{mag}}$ (magenta solid line), so that $B_{0}=10^{14}$G is not enough to satisfy Eq.(\ref{eq:condition eta}) at $31$ ms postbounce.

The radial profiles of potentials at $331$ ms postbounce are shown in the bottom panel of Fig.\ref{fig:potential ECSN}. The potential $\zeta$ is not a dominant term in neutrino Hamiltonian because the neutrino luminositiy on the surface of the PNS decreases as the explosion time has passed. Therefore, in the later explosion phase, the contribution from neutrino-neutrino interactions to the neutrino-antineutrino oscillations are negligible. On the other hand, the matter effect becomes dominant up to $r=631$ km. The shock wave propagates outwards heating material and changing the value of $Y_{e}$ in outer layers of the progenitor model. The vacuum potential $\omega$ is the largest term among neutrino potentials in outer region ($r>631$ km). The decreasing baryon density near the surface of the PNS reduces value of $\lambda$, which results in the crossing of $\eta$ (blue solid line) and $\Omega_{\mathrm{mag}}$ (magenta solid line) in the bottom panel of Fig.\ref{fig:potential ECSN}.

Fig.\ref{fig:B_min radial profile ECSN} represents the radial profiles of Eq.(\ref{eq:minimum initial magnetic field at r}) in ECSN. The minimum value of $B_{0,\mathrm{min}}(r)$ at 31 (331) ms postbounce is $2.16\times10^{14}$($6.62\times10^{13}$) G, respectively. In the early explosion phase (red solid line), the decreasing $B_{0,\mathrm{min}}(r)$ at $r=100-838$ km reflects decreasing $\zeta$, so that neutrino-neutrino interactions suppress the neutrino-antineutrino oscillations even though matter potential $\lambda$ disappears at the outer layer of $\alpha$-elements. The dilute baryon density in the later explosion phase induces small values of $B_{0,\mathrm{min}}(r)$ around $100$ km (green solid line). In a core-collapse scenario to produce a magnetar whose magnetic field is $\sim10^{14}$G, the equilibration of neutrino-antineutrino oscillations would be possible in the later explosion phase.\\

\bibliography{ref}


\end{document}